\documentclass[twocolumn]{aastex631}
\usepackage{amsmath}
\usepackage{hyperref}

\newcommand{\textedit}[1]{#1}
\received{May 14, 2026}
\revised{August 26, 2026}
\accepted{August 31, 2026}

\submitjournal{ApJ}

\shorttitle{Halo Tumbling Induced by Torques from Massive Mergers}
\shortauthors{Ash et al.}

\graphicspath{{./}{figures/}}
\usepackage{printlen}
\usepackage{layouts}

\begin{document}

\title{Dark Matter Halo Tumbling Induced by Torques from Massive Mergers}

\correspondingauthor{Neil Ash}
\email{nfash@umich.edu}

\author[0009-0003-7613-3109]{Neil Ash}
\affiliation{University of Michigan Department of Astronomy \\
1085 S. University \\
Ann Arbor, MI 48109, USA}

\author[0000-0002-6257-2341]{Monica Valluri}
\affiliation{University of Michigan Department of Astronomy \\
1085 S. University \\
Ann Arbor, MI 48109, USA}

\begin{abstract}
Dark Matter halos exhibit slow tumbling known as `figure rotation'. Figure rotation occurs in roughly $80\%$ of isolated $\Lambda$CDM halos and is relatively unaffected by baryonic feedback, making it an attractive and relatively unexplored test for $\Lambda$CDM. However, galaxy mergers are capable of exerting strong torques on a host halo's quadrupole, warranting an investigation of whether such interactions are capable of influencing figure rotation. Using isolated $N-$body simulations and 6 TNG50 halos containing MW analog galaxies, we investigate figure rotation in the context of massive mergers. For isolated galaxies we find that massive mergers elicit rotation of the host halo in both a rapid transient phase during the satellite infall with pattern speeds between $10^\circ-60^\circ$ Gyr$^{-1}$, and post-merger as a steady tumbling with pattern speeds between $2^\circ-40^\circ$ Gyr$^{-1}$ lasting for at least $6-8$ Gyr. Both transient and long-lived tumbling depend on the trajectory of the infalling satellite. We find that figure rotation can be induced by radial mergers with no orbital angular momentum, suggesting that tidal torques from the satellite are a driving mechanism. Meanwhile, TNG50 halos show quadrupole tumbling modified during major mergers in both pattern speed (3/6 halos) and rotation axis orientation (6/6 halos). Quadrupole tumbling remains stable after the last major merger in 5 of our 6 halos. Our results suggest that, rather than tracing the early tidal shear, figure rotation at present day may instead retain a dynamical memory of a halo's last major merger.

\end{abstract}


\section{Introduction} \label{sec:intro}

Dark matter (DM) figure rotation is a phenomenon in which the elongated, `triaxial' DM halos surrounding galaxies steadily tumble over time, in a motion analogous to that of stellar bars. Figure rotation has emerged as a robust prediction of $\Lambda$CDM cosmological simulations, occurring in $80\%-90\%$ of all isolated halos with pattern speeds log-normally distributed from a median of $10^\circ-11^\circ$ Gyr$^{-1}$ and width of $33^\circ$ Gyr$^{-1}$ \citep{bailin_figure_2004,bryan_figure_2007,ash_figure_2023}. Figure rotation may remain stable for several Gyr in isolation \citep{bryan_figure_2007, ash_figure_2023}. Because figure rotation is abundant in $\Lambda$CDM halos and is relatively unaffected by baryonic feedback and adiabatic contraction \citep{ash_figure_2023}, it is a potentially valuable but relatively unexplored test for $\Lambda$CDM.

Figure rotation is capable of imparting observable perturbations in disk galaxies and their stellar halos. \cite{dubinski_warps_2009} find that torques from tumbling halos are capable of inducing warps in their embedded disks, in addition to spiral structure near the disk edge and may even be able to induce the formation of a stellar bar. A possible correlation between figure rotation and disk warping in cosmological halos was recently reported by \cite{johri_disentangling_2026}. Orbits within the halo are also subjected to torques (modeled as Coriolis forces) from the tumbling triaxial figure. The influence of these torques may be measurable in stellar streams \citep{valluri_detecting_2021}. In the context of disk tilting, \cite{nibauer_slant_2024} found that a tumbling quadrupole potential is capable of inducing morphological perturbations to streams dependent on the orientation of the stream's orbital pole with respect to the axis of tumbling, and that failing to account for tumbling in the potential may bias the shape and orientation of the halo inferred by stellar stream orbits. \cite{valluri_detecting_2021} demonstrated it is possible to set an upper limit of $35^\circ$ Gyr$^{-1}$ on the pattern speed of figure rotation in the MW's halo using the Sagittarius stream. 

Despite its prevalence in $\Lambda$CDM simulations, the exact origins of figure rotation are not well understood. \cite{dubinski_cosmological_1992} found that it is possible to induce figure rotation in collapsing halos via tidal torques from large scale structure, resulting in pattern speeds between $\sim7^\circ-90^\circ$ Gyr$^{-1}$. To date, early tidal torquing is the only mechanism that has been demonstrated to be capable of inducing halo figure rotation \citep{dubinski_cosmological_1992}. However, massive satellites are also capable of inducing significant torques on triaxial host halos, especially during close pericentric passages, and are perhaps capable of inducing, modifying, or disrupting halo figure rotation. To the best of our knowledge, the influence of tidal torques from merging satellites on figure rotation has not been explored to date. 

There is reason to expect that major mergers may be significant in inducing or disrupting figure rotation, as they are known to contribute significantly to the total angular momentum content of the halo \citep[e.g.,][]{lagos_quantifying_2018}, have been shown capable of realigning or inducing tilting in stellar disks \citep[][]{dillamore_merger-induced_2022,dodge_dynamics_2023, bell_galaxy_2026}, and have been shown to significantly torque and modify host halo quadrupoles \citep{arora_shaping_2025}. \textedit{\cite{drakos_major_2019} note a steady evolution of the major axis orientation indicative of solid body rotation following a merger between equal mass halos in one idealized simulation.} Given the disruption that the ongoing interaction with the Large Magellanic Cloud is anticipated to produce in our halo \citep[e.g.,][]{garavito-camargo_quantifying_2021,vasiliev_effect_2023,darragh-ford_shaping_2025}, future efforts to constrain or measure figure rotation in the MW's DM halo will first require an enhanced understanding of the role mergers play in inducing or disrupting halo figure rotation. 

This paper is the first in a two-part series investigating the influence of torques from massive companions on dark halo figure rotation. In this paper, we examine the role of massive and primarily ancient mergers using a combination of idealized and cosmological simulations \textedit{in order to answer the following questions:}. 
\begin{enumerate}
    \item \textedit{Are the torques from major mergers alone capable of inducing figure rotation in isolated halos?}
    \item \textedit{If induced by major mergers, do the pattern speeds of figure rotation depend on the orbit of the merging companion?}
    \item \textedit{Do major mergers influence halo tumbling in a cosmological environment in which a halo is subjected to tidal shear from large scale structure and filamentary accretion?}
\end{enumerate}

In paper II, we will look at the response of host halos to LMC-like encounters. The rest of this paper is organized as follows. In Section \ref{sec:simulations} we discuss our idealized and cosmological simulations used. In Section \ref{sec:methods}, we describe our multipole expansion modeling, uncertainty treatment, and quadrupole rotation fitting methods. We present results for both simulation sets in Section \ref{sec:results} and provide discussion and concluding remarks in Section \ref{sec:discussion}.

\section{Simulations} \label{sec:simulations}

To examine the impact of mergers on figure rotation, we make use of both idealized and cosmological simulations. This approach enables us to
\begin{enumerate}
    \item study the impact of massive mergers on a host halo in the absence of confounding factors such as tidal torquing during the initial halo collapse \citep{peebles_origin_1969, dubinski_cosmological_1992}, as well as torques from intersecting filaments and nearby massive companions, and from halo shape evolution due to continual mergers and smooth accretion along filaments \citep{arora_shaping_2025}, and
    \item place the torques due to halo mergers into a cosmological context and assess whether they are still important in light of the confounding effects listed above. 
\end{enumerate}
To this end, we run a suite of idealized simulations in an attempt to marginalize over certain merger properties, and select halos from the TNG50 run from the IllustrisTNG simulation suite. 

\subsection{Idealized simulations}

Our suite of idealized simulations comprises several massive mergers between a triaxial host and a massive satellite galaxy. To keep the parameter space we explore feasible, we restrict our simulations to a single host mass, merger mass ratio, and host halo shape. Instead, we simulate a range of starting points in phase space for our satellite orbit. Further, we use a single Schwarzschild model for our host halo in each of our simulation runs. We do this to ensure that differences in the host's response to the merger between runs reflect the significance of the satellite infall trajectory, and not a potentially different phase space distribution of particles within the host. This allows us to feel confident comparing results between runs, with the drawback that we cannot assess how well our idealized simulation results generalize. We do not consider this a drawback of our approach, since in general it is not possible to reliably extrapolate the results of idealized simulations to a broad population of halos. The goal of these idealized simulations is therefore to build intuition rather than a prescription.

\subsubsection{Density profiles}

We choose a host halo with $M_{200}^\text{host} = 10^{11}$ M$_\odot$ with minor:major and intermediate:major axis ratios of $0.7$ and $0.5$, respectively. This relatively low host mass is used to reduce the computational cost of our simulations while simultaneously retaining a particle mass resolution similar to that used by TNG50. The halo follows the density profile defined by \cite{hernquist_analytical_1990}. We make use of the Planck 2015 \citep{planck_collaboration_planck_2016} cosmology (assuming $\Omega_b=0$, since we neglect baryons) and the mass-concentration relation of \cite{bullock_profiles_2001} to obtain a virial radius $R_{200}^\text{host}=97.9$ kpc and scale radius of $10.5$ kpc. We also place a Gaussian density truncation with a truncation radius at $10\%$ beyond the virial radius, to ensure that orbits with very large apocenters are excluded from our Schwarzschild models. 

Satellites were initialized with a Hernquist profile and scale radius using the \cite{bullock_profiles_2001} mass-concentration relations. The satellites were spherically symmetric rather than triaxial, and were set to have a virial mass $M_{200}^\text{sat}=0.3\cdot M_{200}^\text{host}$, yielding a virial radius $R_{200}^\text{sat}=65.5$ kpc. 

\subsubsection{Initial conditions}

To generate a triaxial $N-$body model for the host halo, we use the Schwarzschild orbit-superposition method \citep{schwarzschild_numerical_1979, schwarzschild_triaxial_1982}. Orbit sampling, integration, and weight determination were performed using the action-based galactic modeling software \texttt{AGAMA} \citep{vasiliev_agama_2019}. We generate an orbit library of $5\times10^{4}$ orbits, and evolve each of these orbits for 50 orbital periods. Weighting coefficients for each of these orbits were determined by each orbit's contribution to a set of 25 shells spaced nearly logarithmically between $0.1$ kpc to $1.1\times R_{200}^\text{host}$, while simultaneously enforcing velocity isotropy. We also enforce that our $N-$body model is initialized with approximately zero net angular momentum. From this orbit library, we produce an $N-$body snapshot by sampling $5\times10^5$ particles. This yields a mass resolution of $2\times10^5$ M$_\odot$. \textedit{We confirm that this model is in equilibrium by evolving it in isolation for 8 Gyr. The density structure of the halo between $3-80$ kpc and the minor:major and intermediate:major axial ratios each fluctuate by $\lesssim1\%$, and the principal axis orientations drift by $\lesssim1^\circ$ over 8 Gyr (see appendix \ref{sec:null_model_stability} for details).}

We produce an equilibrium model for our spherical satellite using the Eddington inversion method \citep{eddington_distribution_1916,binney_galactic_2008} with an isotropic distribution function and zero net angular momentum as implemented with the \texttt{AGAMA} \texttt{GalaxyModel} framework. For each satellite model, we sample a number of particles to match the mass resolution of the host halo. We produce two sets of simulations varying the initial positions and peculiar velocities of the satellites, with the goal of probing the response of the halo under various satellite orbits. We discuss the initial condition choices in more detail in section \ref{sec:idealized_sims} alongside our results.

\subsubsection{Gravity solver and timestepping}

We make use of \texttt{pyfalcon}\footnote{https://github.com/GalacticDynamics-Oxford/pyfalcon} Python interface to the \texttt{falcON} fast multipole gravity solver \citep{dehnen_very_2000} with a leapfrog time-stepping routine to evolve each of our simulations. We choose a Plummer softening kernel with a length of 0.3 kpc, chosen to match the softening used by TNG50. With this choice of softening length, we pick a uniform timestep of $0.98$ Myr, which is similar to the minimum crossing time of the softening length in our $N-$body model. \textedit{Our softening length of 0.3 kpc and particle mass of $2\times10^5$ M$_\odot$ were chosen to approximately match the mass and spatial resolution of TNG50. Literature estimates for the ideal softening length for our system size and particle number range from 0.2 kpc \citep{zhan_optimal_2006} to 0.49 kpc \citep{power_inner_2003} and bracket our choice of 0.3 kpc. We find that our results are insensitive to softening lengths in a range of 0.15-0.6 kpc (see Appendix \ref{sec:resolution_convergence}).}

We evolve each model for $8$ Gyr, with the exception of an individual `fiducial' run which we evolve for $10$ Gyr.

\subsection{IllustrisTNG Simulations}

TNG50 \citep{nelson_illustristng_2019,nelson_first_2019, pillepich_first_2019} is the highest resolution run of the IllustrisTNG simulation suite. TNG50 evolves a volume of $(51.7 \text{ cMpc})^3$ containing $2\times2160^3$ DM + gas particles, yielding a DM mass resolution of $4.5\times10^5$ M$_\odot$ and mean baryon mass resolution of $8.5\times10^4$ M$_\odot$. The stellar and DM softening lengths at present day are each $290$ pc. The simulation is run using the Planck 2015 $\Lambda$CDM cosmology \citep{planck_collaboration_planck_2016} with $h=0.6774$. 

We select halos identified as hosting Milky Way (MW) and M31 analogs by \cite{pillepich_milky_2024}. These analogs are defined as containing a stellar mass of $10^{10.5-11.2}$ M$_\odot$ within $30$kpc, and a stellar disk (minor:major axis ratio $<0.45$) exhibiting spiral arms. Additionally, the host halo must contain a virial mass $M_{200c}<10^{13}$ M$_\odot$ and there may be no neighboring galaxy with a stellar mass $>10^{10.5}$ M$_\odot$ within 500 kpc. These criteria produce a catalog of 198 halos at present day. \cite{pillepich_milky_2024} further identify 6 galaxies within this catalog which they consider to be the closest MW analogs. These galaxies contain thin and thick disk components with heights between $175-360$ and $625-1450$ pc, respectively, disk scale lengths between $1.7-2.6$ kpc, and stellar masses between $10^{10.5-10.9}$ M$_\odot$. The subfind IDs for these six galaxies are 516101, 535774, 538905, 550149, 552581, and 566365. We take these six MW analogs as our primary cosmological halo catalog.

\section{Methods}\label{sec:methods}

\subsection{Multipole Expansion Modeling}\label{sec:methods:expansion_modelling}

We represent the global potential of each halo using the flexible multipole expansion framework as implemented in \texttt{AGAMA}\footnote{https://agama.software/}. Detailed descriptions of the mathematical formalism of multipole expansion models can be found in \cite{binney_galactic_2008} chapter 2.4 and its implementation in the \texttt{AGAMA} documentation (\href{https://arxiv.org/abs/1802.08255}{arxiv:1802.08255}). In brief, this method describes the angular dependence of the potential $\Phi(r,\theta,\phi)$ and/or density $\rho(r,\theta,\phi)$ at fixed radii using the spherical harmonic basis functions:
\begin{align}
    \{\Phi,\rho\}(r,\theta,\phi) &= \sum_{\ell=0}^{\ell_{max}} \sum_{m=-\ell}^{\ell} \{\Phi,\rho\}_{\ell m}(r) Y_\ell^m(\theta,\phi),  
     \label{BFE_basis}
\end{align}
Here, $\ell_{max}$ gives the maximum degree truncating the summation. In practice one may also set a maximum order $m_{max}\leq\ell_{max}$ to limit the number of expansion coefficients, especially for systems with e.g., axial symmetry. Because we consider cosmological halos for which no symmetry is guaranteed, and because we do not attempt to align the reference frame of our expansion to the halo principal axes, we keep the full expansion order with $m_{max}=\ell_{max}$.

For mildly triaxial systems, $\ell_{max}=6$ typically provides a sufficiently high degree to accurately recover the potential. The power contained by a given degree $\ell$ is given by the sum of squared coefficients belonging to that degree:
\begin{equation}
    P_\ell^{\{\Phi,\rho\}} = \sum_{m=-\ell}^\ell \{\Phi,\rho\}_{\ell m}^2.
    \label{eq:coeff_power}
\end{equation}
In eqs. \ref{BFE_basis} and \ref{eq:coeff_power}, we use the notation $\{\Phi,\rho\}$ to denote that the expansion may be done for either the density or potential, and not to represent a product.

Often, basis function expansion models make use of radial basis functions to encode the radial dependence of the density/potential. Rather than radial basis functions, we adopt a multipole model and evaluate the coefficient weights $\rho_{\ell m}(r)$, $\Phi_{\ell m}(r)$ on a set of $N_{grid}$ logarithmically-spaced concentric shells and capture the radial dependence by interpolating the coefficient values between shells using a quintic spline. We choose to set $N_\mathrm{grid} = 50$, with an inner radius of 0.1 $h^{-1}$ kpc which is comparable to the softening length at $z=0$ and an outer radius at the present-day virial radius $R_{200}$.

The multipole expansion coefficients contain useful information on the assembly history and (dis)equilibrium of a given system. The $\ell=0$ monopole term, which is spherically symmetric, encodes the global density profile of the halo. Higher-order terms act as corrections to the monopole which account for deviations from spherical symmetry. For equilibrium or slowly evolving systems, the $\ell=2$ quadrupole terms are the next most significant in the coefficient series, and encode both the orientation and most of the shape of spheroidal or mildly triaxial systems like halos found in simulations \citep[e.g.,][]{chua_shape_2019} as well as stellar bars \citep{ash_stellar_2024}. Odd order coefficients ($\ell=1,3,5,...$) occur with significant amplitude only when an offset is formed between the halo center of mass and the halo centroid, such as when a halo is perturbed by a massive companion during a merger \citep{darragh-ford_shaping_2025,arora_shaping_2025,garavito-camargo_quantifying_2021}. 

\begin{figure*}
    \begin{minipage}[c]{0.63\textwidth}
    \centering
    \includegraphics[width=\textwidth]{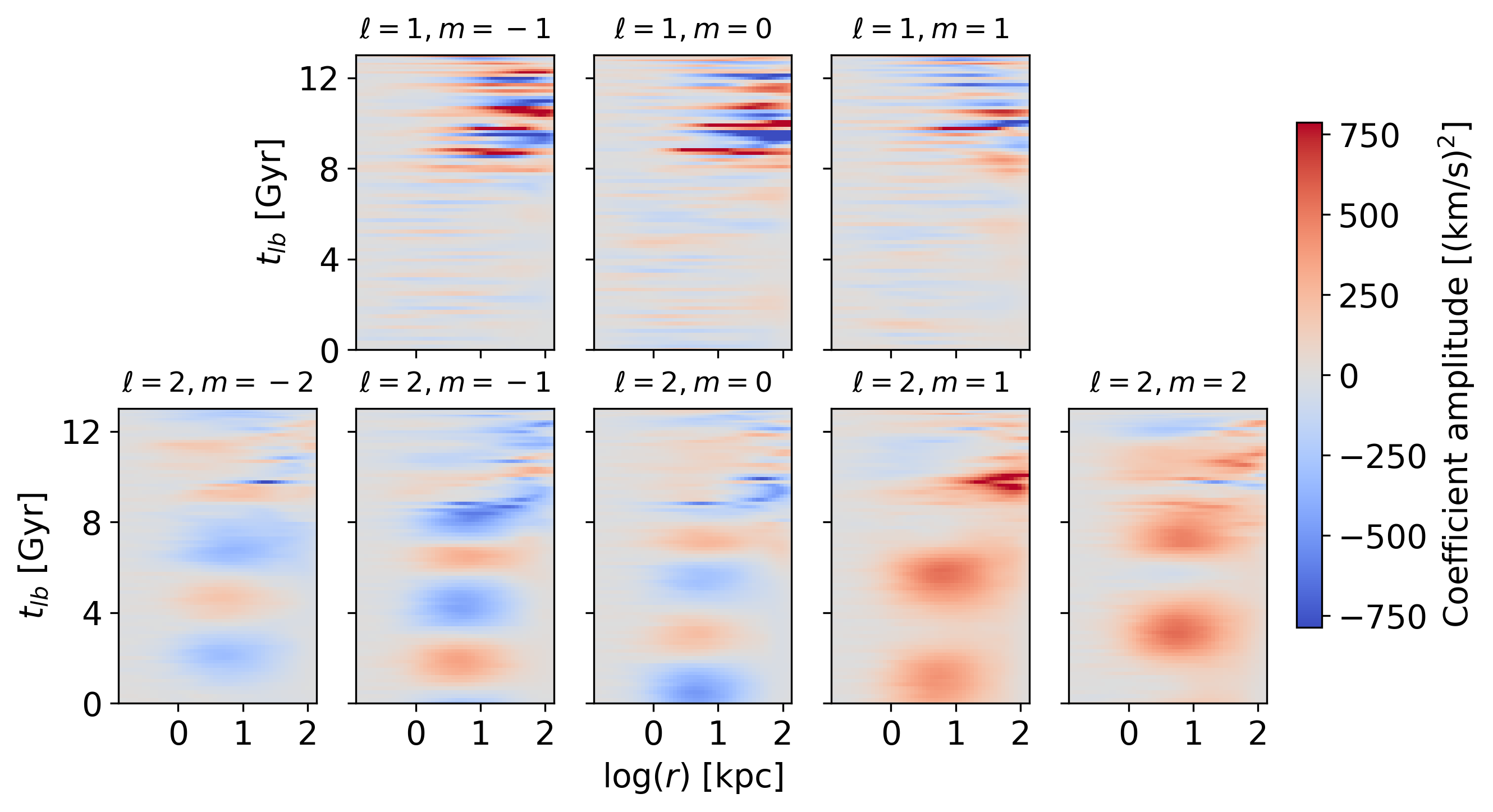}
    \end{minipage}
    \hfill
    \begin{minipage}[c]{0.37\textwidth}
    \includegraphics[width=\textwidth]{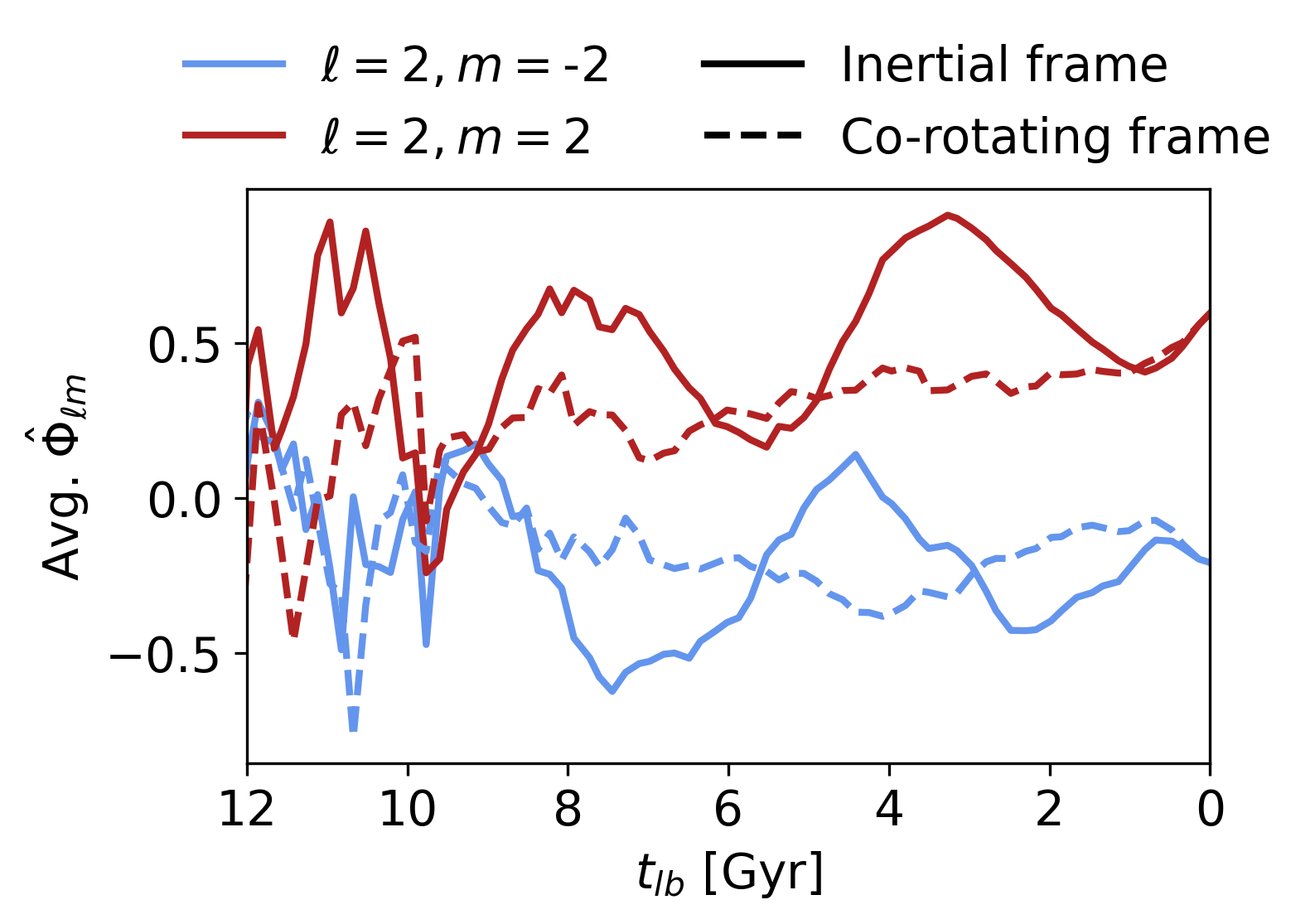}
    \end{minipage}
    \caption{\textit{Left}: Dipole (top row) and quadrupole (bottom row) coefficient amplitudes of the potential against lookback time (vertical axis) and radius (horizontal axis) for a sample halo (GrNr 1425) from the DM only TNG50 run. The presence of figure rotation is made evident in the quadrupole series by the approximately sinusoidal time variation of each of the coefficient amplitudes throughout $-2 \leq \text{log}(r/R_{200}) \leq -0.5$ for $t_{lb}\lesssim 8$ Gyr. The dipole and quadrupole series reveal a series of mergers from $12 \gtrsim t_{lb} \gtrsim 8$ immediately prior to the period of figure rotation. \textit{Right}: $\ell=2,m=\pm2$ normalized quadrupole coefficients for the same halo averaged over radius in the inertial frame and rotated into the co-rotating frame determined by our quadrupole fitting procedure. The oscillations in the inertial frame caused by figure rotation are greatly reduced in co-rotating frame until the time of the mergers. This halo has a particularly visible signature of figure rotation because of its high pattern speed, around $37^\circ$Gyr$^{-1}$.}
    \label{fig:example_l2_figure_rotation}
\end{figure*}

When taken to a sufficiently high order, multipole models are capable of recovering the potential of orbiting substructures within the main halo such as a dynamical friction wake or bound subhalos. In contrast to the main halo, there is no clear physical meaning to the coefficient contribution for a given substructure. We can predict this coefficient contribution, however, by decomposing the potential into its contributions from the host halo and a subhalo, assuming that the substructure is sufficiently compact (see Appendix \ref{sec:substructure_contribution_derivation} for discussion and derivation). When this assumption holds we find that
\begin{equation}
    \rho_{\ell m}^{\text{sub}}(a) = \frac{M_\text{sub}}{a^2}Y_\ell^m(\theta_{\text{sub}},\phi_{\text{sub}})\delta(a-r_\text{sub})
\end{equation}
for the density and 
\begin{equation}
\begin{split}
    \Phi_{\ell m}^{\text{sub}}(a)
= \frac{-4\pi G M_\text{sub}}{r_\text{sub}} \frac{1}{2\ell+1} Y_\ell^{m}(\theta_\text{sub},\phi_\text{sub}) \\ \times
\begin{cases}
 \left(\frac{r_\text{sub}}{a}\right)^{\ell+1} \text{ for }   r_\text{sub} < a \\
 \left(\frac{r_\text{sub}}{a}\right)^{-\ell} \text{ for }   a < r_\text{sub}
\end{cases}
\end{split}
\end{equation}
for the potential, where $(r_{\text{sub}},\theta_{\text{sub}},\phi_{\text{sub}})$ are the galactocentric spherical coordinates of the centroid of the substructure, and $M_{\text{sub}}$ its total mass. Each coefficient in the expansion series picks up some amplitude because the substructure is offset from the origin. The coefficient power spectrum contributed by a substructure evidently scales as 
\begin{equation}
\begin{split}
    P_{\ell}^{\Phi, \text{sub}}(a) \propto \left(\frac{M_{\text{sub}}}{r_{\text{sub}}}\right)^2(2\ell+1)^{-2} \\
    \times
\begin{cases}
 \left(\frac{r_\text{sub}}{a}\right)^{2(\ell+1)} \text{ for }   r_\text{sub} < a \\
 \left(\frac{r_\text{sub}}{a}\right)^{-2\ell} \text{ for }   a < r_\text{sub}
 \end{cases}
    \label{eq:subhalo_power_series}
\end{split}
\end{equation}
for the potential and is constant with respect to $\ell$ in density. The contributions of the substructure to each series are highly localized to the radial region that contains the substructure, making the presence of the substructure easily identifiable.

To generate each potential model for a given snapshot in our TNG50 halos, we select all DM particles belonging to the central Subfind subhalo identified by TNG. This choice excludes bound substructures when they can still be tracked by the Subfind algorithm, however other halo substructures not tracked by Subfind will still be included in our expansions. These may include e.g., tidally stripped debris or the remnant cores of stripped subhalos after they are no longer resolved by Subfind. We place the center for the expansions on the particle with the lowest potential, which closely matches the halo's minimum potential.  

In our idealized simulations, we produce our multipole expansion models using all particles originally belonging to the host halo. Rather than the potential minimum, we identify a halo centroid in our idealized simulations using an iterative approach by finding the mean particle position in a sphere of radius $20$ kpc centered on the current centroid estimate. We then update our centroid estimate and repeat this procedure until convergence. We use the multipole models of our idealized simulations only for fitting the quadrupole rotation, and find that identifying the centroid as the potential minimum or using this iterative method does not affect our results.

\subsection{Coefficient transformation under rotations}

Conveniently, the rotation $\mathbf{R}$ of any spherical harmonic function $Y_\ell^m$ can be expressed by a linear combination of other spherical harmonic functions with the same degree $\ell$:
\begin{equation}
    Y_m^\ell(\mathbf{R}^{-1}\mathbf{\Omega}) = \sum_{m'=-\ell}^{\ell} \mathfrak{D}^{(\ell)}_{mm'}[\mathbf{R}]Y^\ell_{m'}(\mathbf{\Omega}).
    \label{eq:SPH_rotation}
\end{equation}
The weighting coefficients $\mathfrak{D}^{(\ell)}_{mm'}$ are elements of the Wigner $\mathfrak{D}$-matrices, defined as
\begin{equation}
\begin{split}
        \mathfrak{D}^{(\ell)}_{mm'}[\mathbf{R}] &\equiv \left<\ell m|\mathbf{R}|\ell m'\right> \\
        &= \oint d^2\mathbf{\Omega} Y_m^{\ell*}(\mathbf{R\Omega})Y_{m'}^\ell(\mathbf{\Omega}).
\end{split} \label{eq:D-matrix_elements}
\end{equation}

The rotation applied by the $\mathfrak{D}$-matrices can be viewed equivalently as the rotation of the spherical harmonic functions (as in eq. \ref{eq:SPH_rotation}), or as the (inverse) rotation of the reference frame in which the harmonics are evaluated. This means that the transformation of the expansion coefficients $\Phi_{\ell m}, \rho_{\ell m}$ for a rotation $\mathbf{R}$ is done in the same manner using the $\mathfrak{D}$-matrices:
\begin{equation}
\mathbf{R}\{\Phi,\rho\}_{\ell m}(r) \equiv \sum_{m'=-\ell}^\ell \mathfrak{D}^{(\ell)}_{mm'}[\mathbf{R}]\{\Phi,\rho\}_{\ell {m'}}(r),
    \label{eq:coeff_rotation}
\end{equation}
where $\mathbf{R}\{\Phi,\rho\}_{\ell m}$ is shorthand notation for the rotation $\mathbf{R}$ applied to coefficients. At a given $\ell$, the corresponding Wigner $\mathfrak{D}$-matrix is a unitary matrix of shape $(2\ell+1)\times(2\ell+1)$, which also gives these transformations the property of conserving the power associated with the degree $\ell$, as given by equation \ref{eq:coeff_power}. 

\subsection{Quadrupole rotation fitting}\label{sec:methods_rotation_fitting}

Halo figure rotation manifests as a fixed triaxial halo shape which rotates steadily in time about some axis. With its triaxial shape fixed and in the absence of any significant perturbers, the power expressed in the halos quadrupole $P_{\ell=2}^\Phi$ will also remain fixed in time, as will the values of its quadrupole coefficients $\Phi_{\ell m}(r)$ and $\rho_{\ell m} (r)$ in the reference frame co-rotating with the halo. In the inertial reference frame, the quadrupole coefficients will show semi-sinusoidal periodic oscillations as the corresponding basis functions come in-to and out-of phase with the triaxial figure. See figure \ref{fig:example_l2_figure_rotation} for an example of these oscillations in the quadrupole coefficients of a sample halo undergoing figure rotation, as well as the averaged $m=\pm2$ quadrupole coefficients in the inertial frame and rotated into the co-rotating frame. 

We use the quadrupole rotation fitting method\footnote{Reasonable interpretation is required to assess whether the tumbling we measure is indicative of figure rotation or orbiting substructures, and to make this distinction clear we refer to our fit results as `quadrupole tumbling'.} presented by \cite{ash_stellar_2024} to measure the rotation of stellar bars from their multipole coefficients, and provide a summary of the method here. Conceptually, by correctly rotating the quadrupole harmonic coefficients into the co-rotating frame, the time evolution associated with figure rotation can be minimized. This property is used to define a $\chi^2$, which measures the time variance of the quadrupole coefficients between a set of snapshots ($t_j$) and the set of their previous snapshots ($t_{j-1}$):
\begin{align}
 &\chi^2(\mathbf{R}) = \label{eq:chi2_rotation} \\
 &\left. \frac{1}{\sum_j w_j} \sum_{i,j,m}\frac{w_j\{\hat{\Phi}_{\ell m}(t_{j-1},r_i)-\mathbf{R}\hat{\Phi}_{\ell m}(t_j,r_i)\}^2}{\sigma^2[\hat{\Phi}_{\ell m}(t_j,r_i)]}\right|_{\ell=2}, \nonumber
 \end{align} 
where the $\chi^2$ summed over a set of snapshots $t_j$, radii $r_i$, and the five quadrupole coefficients $-2\leq m \leq 2$. The $\hat{\Phi}_{\ell,m}$ are the multipole expansion coefficients normalized by the power contained in the $\ell$ coefficient set:
\begin{equation}
    \hat{\Phi}_{\ell m} = \frac{\Phi_{\ell m}}{\sqrt{P^\Phi_\ell}}.
\end{equation}
We perform this normalization to remove the time variation in the coefficients associated with changes in the quadrupole strength, and to improve the numerical stability of our fitting.  $\mathbf{R}\hat{\Phi}_{\ell m}(t_j,r_i)$ is a set of rotated coefficients as defined in eq \ref{eq:coeff_rotation}, $w_j$ is a set of optional weighting coefficients, and $\sigma^2[\hat{\Phi}_{\ell m}(t_j,r_i)]$ is a bootstrap coefficient uncertainty, determined by resampling particles when generating coefficients. Alternatively, one could normalize the quadrupole series by the monopole coefficients $\Phi_{00}$; we do not adopt this normalization because we found it placed too much weight in the outer halo, where the quadrupole orientation is more uncertain and contributes only marginally to the global quadrupole potential, whose tumbling we seek to characterize.

The $\hat{\Phi}_{\ell m}$ are each measured in the inertial frame; the rotation $\mathbf{R}$ attempts to match the time evolution caused by changes in quadrupole orientation. We specify $\mathbf{R}$ with three parameters; the orientation of the rotation axis $(\phi_\mathbf{R},\theta_\mathbf{R})$, and the pattern speed $\Omega_p$. The angle of rotation between snapshots $j-1$ and $j$ is given by $\Omega_p\Delta t$, where $\Delta t$ is the (possibly nonuniform) snapshot spacing. We note that with this approach, we may fit several snapshots simultaneously with a single fixed pattern speed, even if the snapshots are not uniformly spaced in time. 

Conceptually, the rotation $\mathbf{R}$ that minimizes this $\chi^2$ gives the transformation into the co-rotating frame, in which the time-variation of each individual quadrupole coefficient is minimized. While we are interested in the tumbling of the quadrupole specifically, this methodology is generic and could be applied to other coefficient degrees, such as the dipole, or instead to several coefficient degrees simultaneously.

When applying this method, we attempt to measure the instantaneous quadrupole rotation rate by performing a fit centered on each individual snapshot. We select a window of 30 snapshots to use in the fit with Gaussian weights $w_j$ centered on the current snapshot and with a width $\sigma_w = 1.5 $ Gyr. The choice of the Gaussian kernel and window length influence the output fit; a wider kernel will capture long-lived rotation but will necessarily make the fit less sensitive to rotations occurring over short timescales, such as may occur during the pericenter passage of a massive satellite. Smaller kernels are able to capture these transient signatures, but produce noisy measurements of long-lived quadrupole rotation. We choose a width of $1.5$ Gyr to match the expected durations for figure rotation observed in \cite{ash_figure_2023}, and to prioritize the recovery of rotation that is stable over this timescale. 

To exclude regions of the halo where the shape may be dominated by adiabatic contraction due to the disk, we omit radii from our fit for which the enclosed baryonic mass exceeds the enclosed DM mass. This sets our lower limit on the set of $r_i$ radial nodes used in our rotation fitting. The upper limit we set to be at $60\%$ of the virial radius at $z=0$, matching that used by \cite{bailin_figure_2004} and \cite{ash_figure_2023}. With our set of radial nodes, this gives us $24-26$ nodes for fitting in each halo. 

To calculate the Wigner $\mathcal{D}-$matrices and the related coefficient transformations, we use the \texttt{spherical} software package \citep{boyle_spherical_2023}. We additionally represent our rotations as quaternions, handled computationally by the \texttt{quaternionic} package \citep{boyle_quaternionic_2024}.

\subsection{Uncertainty Treatment}

Accurately estimating the uncertainty on the multipole expansion coefficients measured from $N-$body systems is a difficult task, and to the authors' knowledge a complete and precise understanding of these uncertainties has not yet been established. One known error source stems from uncertainties in the determination of the halo centroid; if the halo is not properly centered during expansion some amount of error may be added to each term, affecting in particular the expansion coefficients of the inner halo (this can be inferred from our discussion of substructures in section \ref{sec:methods:expansion_modelling}). The dominant uncertainty for $N-$body systems is often assumed to be Poisson noise, likely because they are sometimes interpreted as discrete observations drawn from a smooth density function. However, Poisson noise alone may underestimate the uncertainty on the density as the positions of particles, and hence their errors, are correlated from mutual interactions under gravity \citep[e.g.,][]{bailin_figure_2004}. 

\begin{figure}
    \centering
    \includegraphics[width=.75\linewidth]{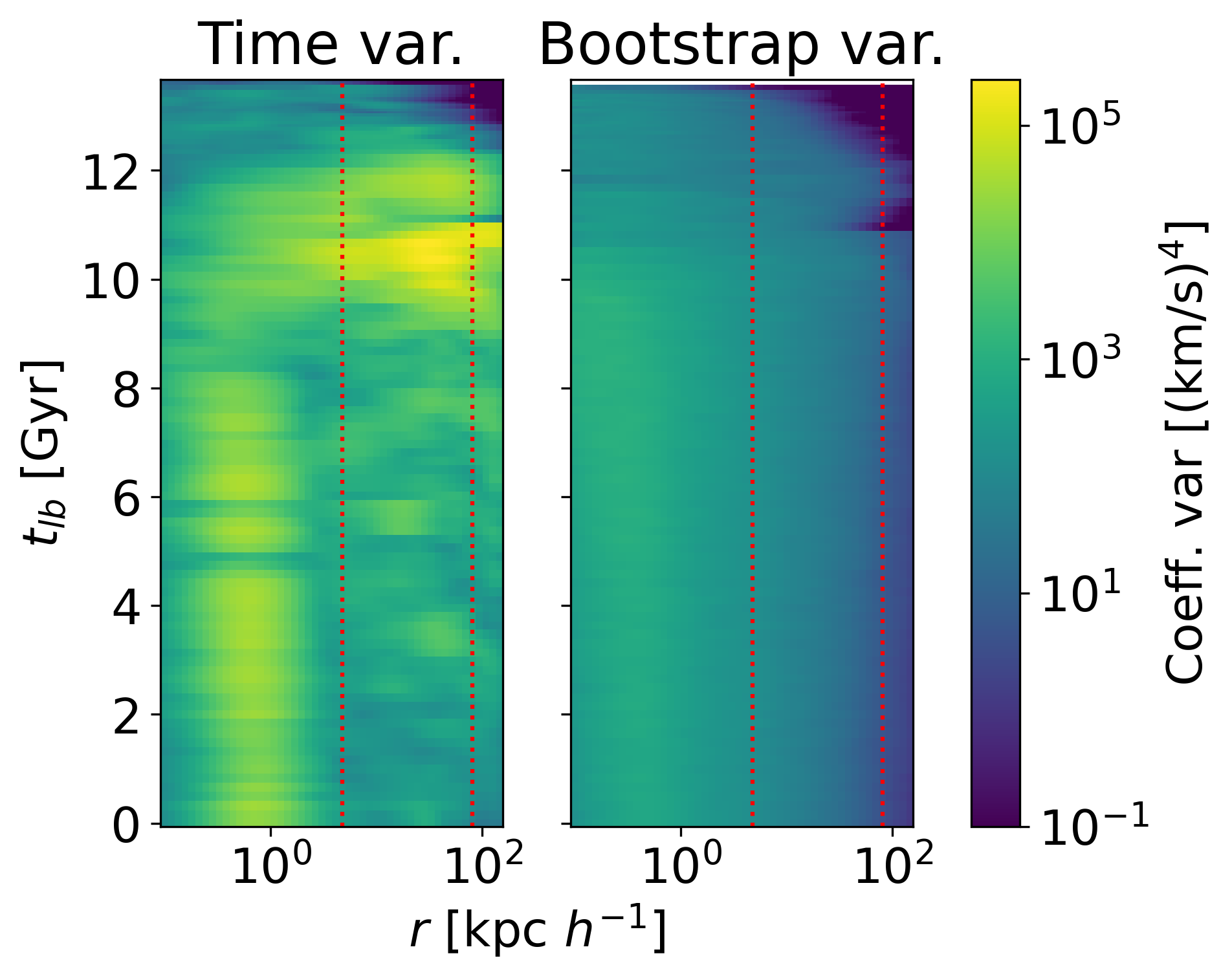}
    \caption{Mean DM quadrupole coefficient variances inferred by bootstrap resampling of halo particles (right) compared to the variance in time (left) against lookback time (vertical axis) and radius (horizontal axis) for TNG50 subhalo 535410. The left panel is measured using a window of three snapshots. Red dotted lines enclose the region used for rotation fitting. Even when this halo is relatively quiescent ($6\gtrsim t_{lb}\gtrsim 0$), the spread in coefficient values between neighboring snapshots in the halo is $1-2$ dex greater than bootstrap uncertainty estimates would suggest. The large time variance around 1 kpc $h^{-1}$ is caused by the stellar bar, while the time variance in the outer halo is likely caused by orbiting substructures.}
    \label{fig:GrNr_211_bootstrap_and_time_std}
\end{figure}

As a fiducial estimate of the uncertainty on our coefficients, we use bootstrap re-sampling of the halo particles used in the expansion. We produce a set of 100 bootstrap samples, allowing replacement, and measure the standard deviation of each of the coefficients at each radial bin for both the density and potential series. However, we note that the time series often show considerably greater variance between snapshots than the bootstrap errors would suggest, even for relatively quiescent halos evolving only slowly in time. We show this effect in one sample system in figure \ref{fig:GrNr_211_bootstrap_and_time_std}; the variance inferred from bootstrapping (right panel) is typically $1-2$ dex lower than the variance of the coefficients in time, measured using a three-snapshot window (left panel). This is most likely caused by correlated errors from orbiting substructures in the halo, which causes bootstrapping to underestimate the coefficient variance. For this reason, we use the bootstrap variance to approximate $\sigma^2[\Phi_{\ell m}(t_j,r_i)]$ used for our rotation fitting (eq. \ref{eq:chi2_rotation}), but estimate our fit uncertainties by performing bootstrap resampling on the snapshots involved in the fit, effectively modifying the snapshot weights $w_j$ according to the number of times a snapshot appears in a given bootstrap sample. In practice we find that this bootstrap fit uncertainty is the most conservative estimate on our fit pattern speeds, and adopt it as our fit uncertainty.

\section{Results}\label{sec:results}

\subsection{Idealized Simulations}\label{sec:idealized_sims}

\begin{figure*}
    \centering
    \includegraphics[width=0.65\textwidth]{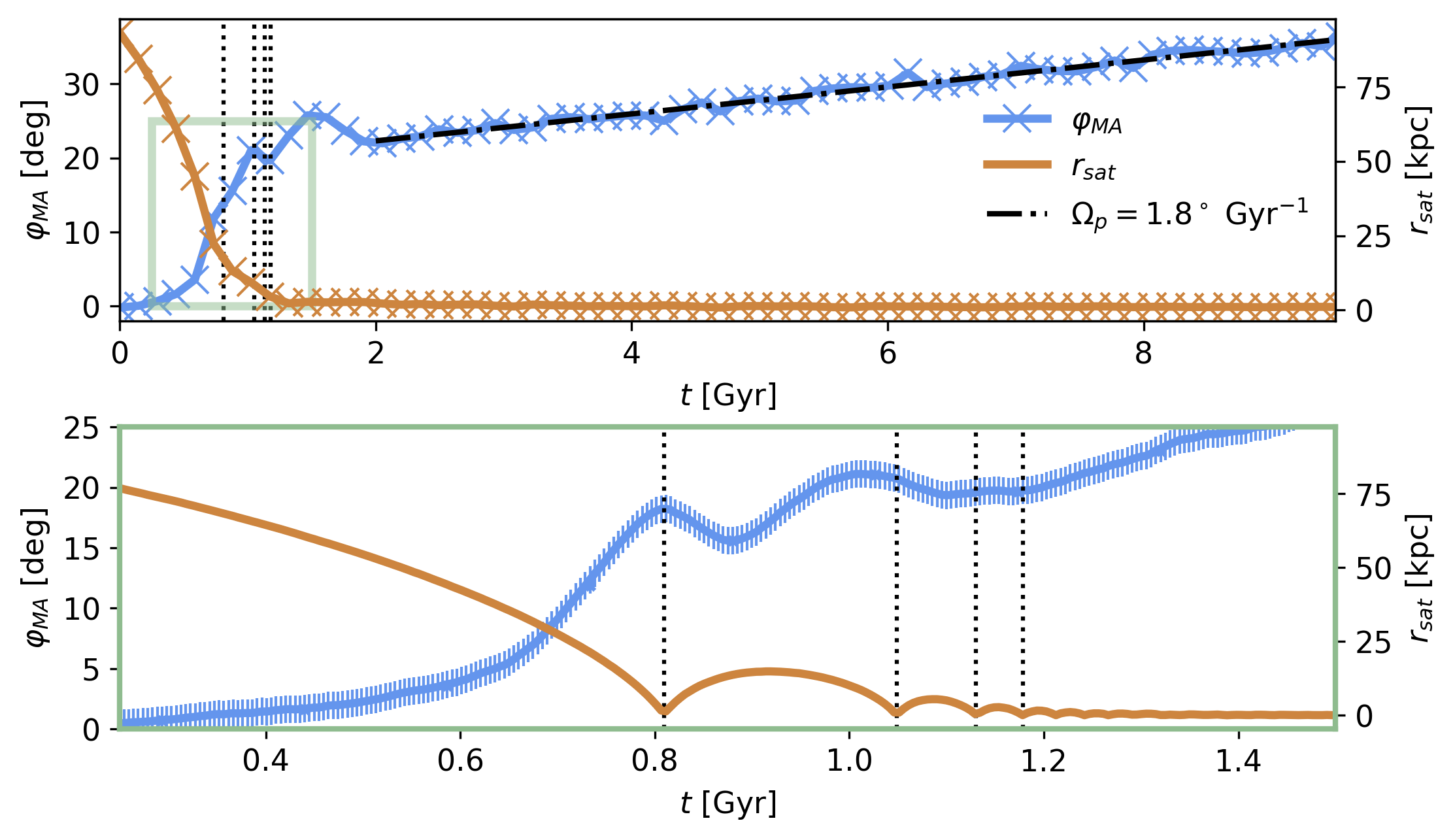}
    \caption{Evolution of the host halo major axis position angle $\varphi_{MA}$ (blue line) for a 1:3 radial merger infalling at $45^\circ$ from the halo major axis. For reference, we plot also the satellite remnant's galactocentric radius (orange line) on the right axis, with the first four pericentric passages marked (vertical dotted lines). The halo major axis undergoes both a transient and long-lived rotation as a result of the satellite merger. In the transient response (see zoom-in in the bottom panel), the major axis swings quickly in response to the satellite during its infall. Post-merger, the halo continues rotating with a stable pattern speed of $1.8^\circ$ Gyr$^{-1}$ for the duration of the simulation. $\varphi_{MA}$ is measured using only particles originally belonging to the host halo.}
    \label{fig:Model_9_phi_transient_and_longlived_response}
\end{figure*}

Our set of idealized simulations test, in a controlled setting, how massive mergers re-orient the quadrupole and major axis of the host halo both as the merger proceeds and long after the merger has concluded. We test the significance of the satellite orbit using two sets of simulations; one set varying the starting position of satellites on a radial infall, and one set varying the velocity direction of satellites beginning at a fixed position. We additionally present one `fiducial' simulation, with a higher snapshot cadence for the first 4 Gyr, of a radial merger infalling $45^\circ$ off the halo major axis, in the plane containing the major and intermediate axis. The set of radial models was chosen to test whether tidal torques from the satellite on the host halo figure are capable of inducing long term halo tumbling in the absence of additional angular momentum brought in by the satellite. The second set of models varying the initial satellite orbital eccentricity allows us to compare the relative importance of tidal torquing and angular momentum transfer from the satellite. We present the major findings from our simulation sets in the following subsections.

\begin{figure*}
    \centering
    \includegraphics[width=0.75\textwidth]{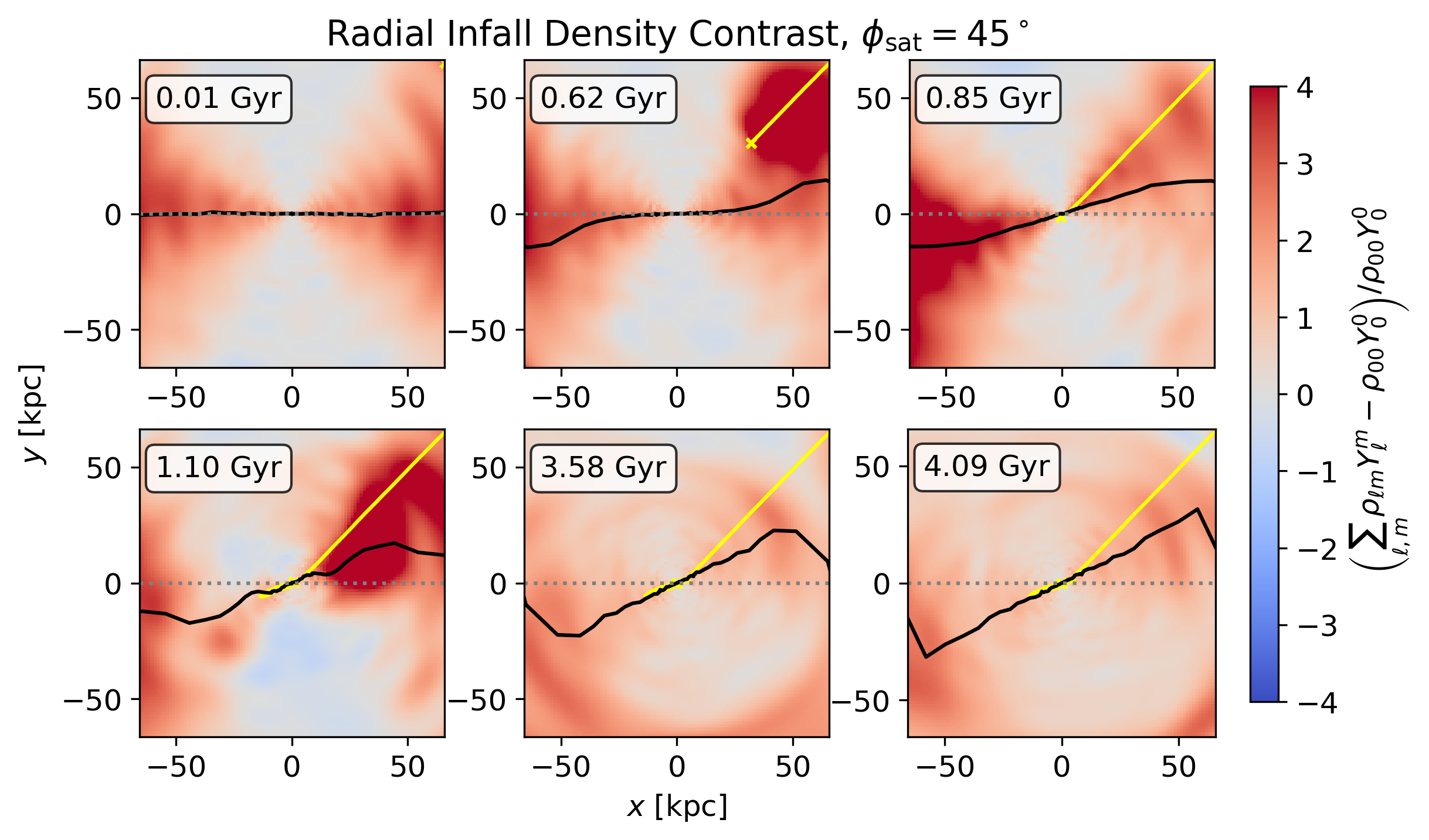}
    \caption{Evolution of the host's midplane density contrast for a 1:3 radial halo merger. The yellow line shows the position and past trajectory of the merging satellite, while the black line shows the phase angle of the plane-projected $m=2$ Fourier density component as a function of radius and time and serves as a proxy for quadrupole orientation. Color shows the ratio of the multipole density to the monopole density in the plane $z=0$, calculated with $\ell_{\max}=10$ and using only the host particles.}
    \label{fig:aspherical_density_timeseries}
\end{figure*}

\subsubsection{Transient and long-lived responses}

In our fiducial idealized simulation, we observe rotation in the orientation of the host's major axis (MA) and quadrupole orientation as both a transient and long-lived response to the satellite merger (see figure \ref{fig:Model_9_phi_transient_and_longlived_response}). The transient response begins during the early infall, when the satellite is still $50-75$ kpc from the host (satellite galactocentric distance is shown on the right-hand $y-$axis in figure \ref{fig:Model_9_phi_transient_and_longlived_response}). The tilting rate of the MA accelerates as the satellite approaches its first pericenter passage, at which time the MA has been displaced by $\approx20^\circ$ from its initial orientation. We show the transient response and satellite galactocentric distance for the same simulation in greater detail with a higher snapshot cadence in the lower panel of figure \ref{fig:Model_9_phi_transient_and_longlived_response}. 

The long-lived response becomes apparent beginning around $t=2$ Gyr, at which point the merger has concluded and the halo major axis continues to smoothly rotate in time about the halo minor axis, with a pattern speed of approximately $1.8^\circ$ Gyr$^{-1}$ (0.03 km s$^{-1}$ kpc$^{-1}$). We emphasize that this response was elicited from a radial merger; the satellite fell onto the host halo initially without any orbital angular momentum. During its infall, the satellite exerts a tidal torque on the host halo's quadrupole, which is capable of rotating the orientation of the quadrupole. The halo exerts an equal and opposing torque on the infalling satellite, causing it to gain orbital angular momentum. 

The above picture describes the scenario well in the limit that the host halo acts as a rigid body, but fails to capture the significance of disequilibrium dynamics within the halo. To better understand the host halo's response to this merger, we examine the aspherical density component, recovered by multipole expansion of the host's material out to $\ell_{\max}=10$. In figure \ref{fig:aspherical_density_timeseries} we show the mid-plane density contrast (i.e. the fractional deviation from the spherically averaged density) across six snapshots from our simulation. We additionally track the satellite's infall trajectory (yellow line) and the plane-projected quadrupole orientation (black line) as a function of radius. 

At the start of the simulation, the aspherical component of the host halo's density is clearly dominated by its quadrupole term as expected (top-left panel of fig. \ref{fig:aspherical_density_timeseries}). During the infall of the satellite, the dynamical friction wake begins to dominate over the quadrupole, and locally shifts the orientation of the quadrupole (top-middle panel). The quadrupole orientation is also shifted in the opposite region of the sky due to a `collective response' type reaction \citep[see e.g.,][]{garavito-camargo_quantifying_2021} of the halo: as the potential minimum/ halo centroid accelerates towards the infalling satellite it leaves behind the line of higher density halo material (i.e., the original major axis), producing an apparent rotation in the major axis orientation of the outer halo (top-right panel). 

An alternative way to see the effect of the collective response term in reorienting the quadrupole comes from our discussion of substructure in section \ref{sec:methods} and Appendix \ref{sec:substructure_contribution_derivation}. The collective response term occurs as an offset of the center of mass of the outer halo from the halo centroid/ potential minimum. While the dominant contribution of the collective response (and indeed any substructure offset from the origin) is to the dipole term, there is a contribution to every degree of the expansion. The contribution of the collective response is evidently capable of significantly reorienting the halo quadrupole. 

The net effect of the combination of 1) the dynamical friction wake, 2) the collective response, and 3) tidal torquing from the satellite is to excite a quadrupolar density wave in the outer halo that propagates inward during the satellite infall. At the time of the first pericenter passage ($t=0.85$ Gyr, top-right panel in fig. \ref{fig:aspherical_density_timeseries}), this density wave reflects through the halo centroid and begins to propagate outwards. The satellite will torque the central halo to align with the direction of its second infall (lower left panel), and the combined torques of the satellite and inner halo create further density waves that begin propagating outwards. Physically, these density waves likely correspond to shells of host halo material which were caught in the gravitational wake of the satellite and lost their angular momentum support. The propagation mechanism for the density waves is gravitational torquing between neighboring shells within the halo. Beginning around $t=1.7$ Gyr, one of these density waves begins to wind up, and becomes a near-standing wave which slowly moves outwards and endures past $t=4$ Gyr (see the pronounced twist in the black line in the lower middle and lower right panels of fig. \ref{fig:aspherical_density_timeseries}). The halo quadrupole becomes dominant again quickly post-merger ($t\approx1.6$ Gyr), and is rotated by approximately 20$^\circ$ with respect to its initial orientation. 

\subsubsection{Infall trajectory dependence}\label{sec:radial_infall}

\begin{figure}
    \centering
    \includegraphics[width=1\linewidth]{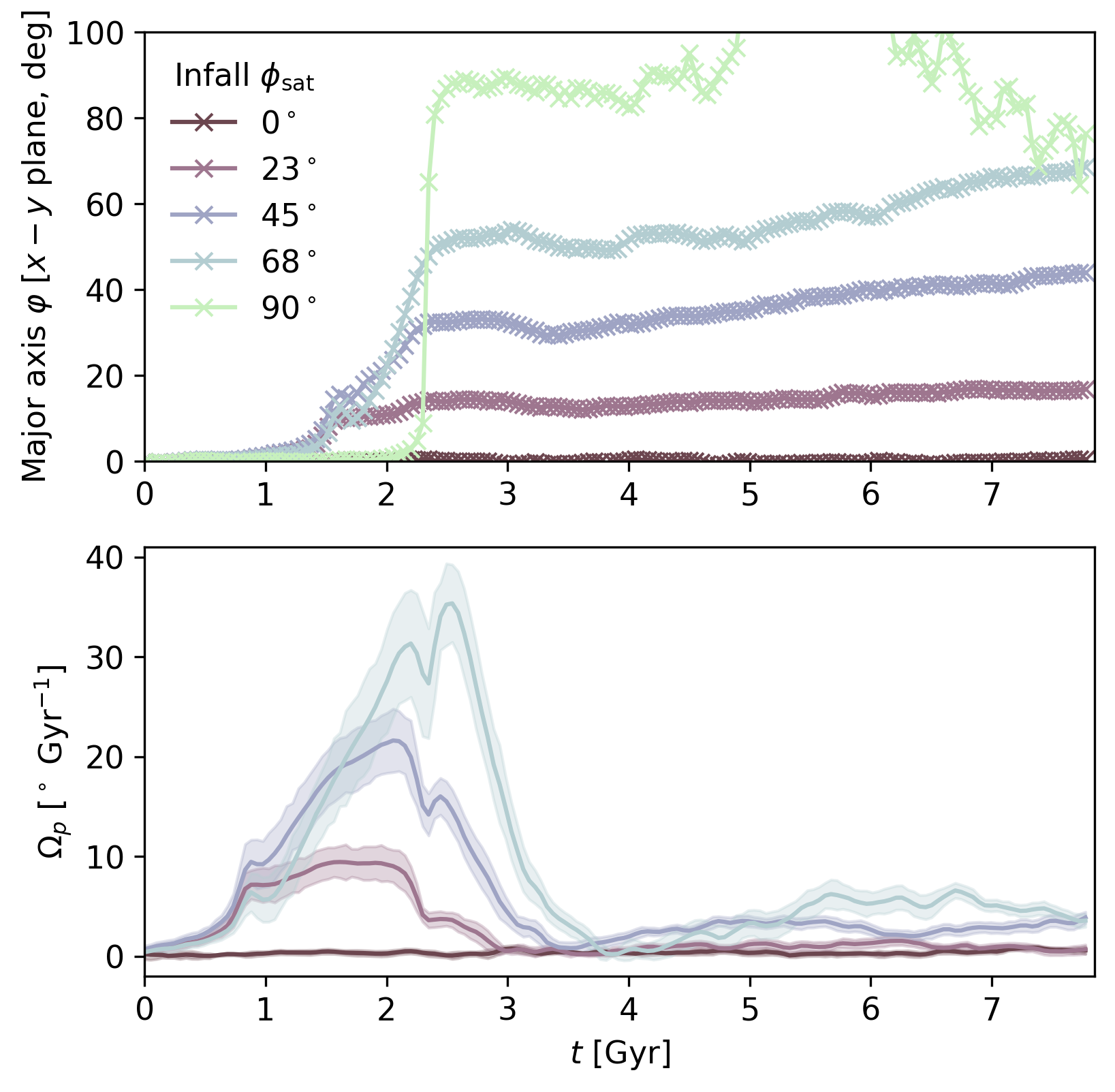}
    \caption{(Top) Response of the host major axis orientation $\varphi(t)$ as a function of satellite infall trajectory in the plane containing the major and intermediate axes. The slope $d\varphi/dt$ corresponds to the pattern speed of figure rotation about the halo minor ($z$) axis. (Bottom) Quadrupole rotation pattern speed $\Omega_p(t)$, fit using the multipole expansion coefficients and the methodology presented in section \ref{sec:methods_rotation_fitting}. We remove the pattern speed fit for the model with $\phi_{\text{sat}}=90^\circ$ due to the large uncertainties as the halo becomes oblate post-merger. The post-merger figure rotation induced in these halos shows pattern speeds which depend on the satellite infall trajectory. }
    \label{fig:Model_10_phi_host_x_y_plane_infall}
\end{figure}

Figure rotation induced by an infalling satellite may encode information on the satellite's past orbital trajectory. To probe the sensitivity of resultant figure rotation to a satellite's infall trajectory, we run a set of five variations of our 1:3 merger simulation, with each satellite undergoing a radial infall at five angles with respect to the orientation of the host halo's major axis, linearly spaced from 0$^\circ$ (i.e., infalling along the halo major axis) to 90$^\circ$ (infalling along the halo intermediate axis). Each of these satellites are initialized within the plane containing the halo's major and intermediate axes, such that their tidal torque on the host halo's quadrupole is aligned with the halo minor axis. Additionally, the satellites are all initialized at a galactocentric distance of 138 kpc (such that virial edges of the two halos are just touching) and given an initial (radial) velocity of approximately $80\%$ the spherically-averaged escape velocity (between $57.7-58.6$ km s$^{-1}$, chosen so that the total energy of the satellite is constant between models). These choices approximate the binding energy of the satellite beginning at rest at 455 kpc.

In the top panel of figure \ref{fig:Model_10_phi_host_x_y_plane_infall}, we plot the evolution of the position angle $\varphi$ of the host's major axis over the duration of the simulation. In this frame, the major axis initially points towards the $x-$axis, and the intermediate axis initially points towards the $y-$axis. The satellite's orbit is mostly confined to the $x-y$ plane.  Evolution of $\varphi$ indicates figure rotation of the halo about its minor axis, with the time derivative $d\varphi/dt=\Omega_p$, the pattern speed of figure rotation. 

If the figure rotation we observe in our idealized simulations is caused by tidal torquing on the host's triaxial halo, then we may expect the highest pattern speed of figure rotation $\Omega_p$ when $\phi_{\text{sat}}\approx45^\circ$, i.e., where the classical quadrupolar torque is maximized.

Two of the five models simulate a satellite falling onto the host halo along either the major ($\phi_{\text{sat}}=0^\circ$) or intermediate ($\phi_{\text{sat}}=90^\circ$) axes. Because the satellites fall onto the halo along an axis of symmetry, they are not capable of exerting a net torque on the host halo's quadrupole. Indeed, we observe essentially no rotation for $\phi_{\text{sat}}=0^\circ$, and our figure rotation fits to this model are consistent with no rotation (bottom panel of figure \ref{fig:Model_10_phi_host_x_y_plane_infall}). For $\phi_{\text{sat}}=90^\circ$, we observe $\varphi$ rapidly evolve from $0^\circ$ to $90^\circ$ just after 2 Gyr, when the merger concludes. This evolution occurs as the original intermediate axis is elongated by the merger, ultimately becoming marginally longer than the original major axis and is therefore not indicative of a steady rotation in the major axis. After the merger has concluded, the halo gradually evolves from triaxial to oblate (intermediate:major axis ratios evolve to $>0.98$ for $t\geq4.6$ Gyr), causing artificial evolution in $\varphi(t)$. This evolution is observed because the orientation of the major axis is not well defined in an oblate halo. The fit uncertainties for this model also become quite large because of its nearly oblate shape and so we remove it from the bottom panel of figure \ref{fig:Model_10_phi_host_x_y_plane_infall} for clarity, although we note that the fit pattern speed remains consistent with zero. Both of these two control models show no rotation post merger, demonstrating that our idealized simulations do not exhibit figure rotation post-merger when the satellite falls in radially along a principal axis, since in such an infall the merging satellite does not induce any tidal torques on the host's triaxial figure. 

The three models in which the satellite fell onto the host halo with an angle $23^\circ \leq \phi_{\text{sat}}\leq 68^\circ$ each show both the transient and long-lived rotation we observed previously (fig. \ref{fig:Model_9_phi_transient_and_longlived_response}). The responses of these three models during the transient response leading up to the first pericenter passage (all models fall within $1.58\lesssim t_\mathrm{peri} \lesssim 1.61$ Gyr) are relatively similar, with the major axes swinging from $10^\circ-15^\circ$ with the greatest response in the model whose $\phi_{\text{sat}}=45^\circ$. By the end of the merger, the major axis of each model has evolved to align with the secondary infall direction of the merging satellite, and it is this alignment to the satellite debris that drives the rotation pattern speed from the first pericenter until the merger is resolved. After the merger is concluded, each of these models exhibits slower figure rotation which is long-lived and has a relatively constant pattern speed for the remainder of the simulation ($\sim5$ Gyr). This long-lived figure rotation also shows a dependence on the merger's orbital parameters; for $3.25\leq t \leq 5$ Gyr the model with $\phi_{\text{sat}}=45^\circ$ tumbles with a pattern speed of $3.6^\circ$ Gyr$^{-1}$, modestly faster than the models with $\phi_{\text{sat}}=23^\circ$ ($1.3^\circ$ Gyr$^{-1}$) or $\phi_{\text{sat}}=68^\circ$ ($1.7^\circ$ Gyr$^{-1}$). These pattern speeds are determined by a linear fit to $\varphi(t)$ and are each consistent with the time-dependent fits to the quadrupole coefficients (lower panel of figure \ref{fig:Model_10_phi_host_x_y_plane_infall}) at various points in this time window. This dependence on the infall trajectory of the satellite implies that the halo retains some dynamical `memory' of the merger for several Gyr in the form of figure rotation. Furthermore, halo tumbling is evidently not induced by orbital angular momentum donated by the satellite during the merger but rather can be initiated purely by tidal torques from a massive, merging satellite.

\subsubsection{Satellite impact parameter}\label{sec:impact_parameter}

\begin{figure}
    \centering
    \includegraphics[width=1\linewidth]{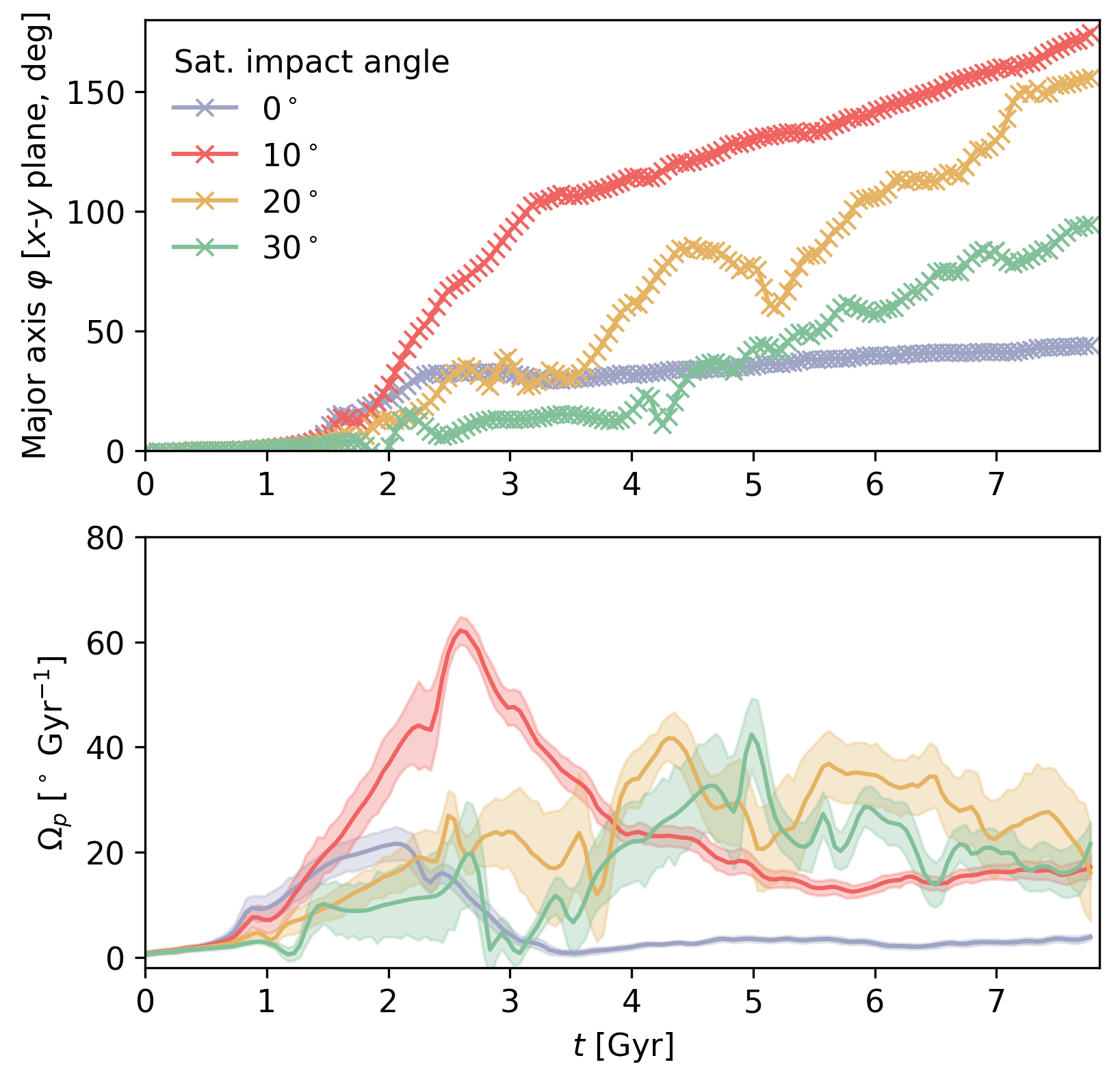}
    \caption{(Top) Response of the host major axis orientation $\varphi(t)$ as a function of the satellite orbital impact angle\. The angles in the legend represent the offset of the satellite's initial velocity vector from a radial infall. Each satellite begins at a position $45^\circ$ above the halo major axis in the plane containing the major and intermediate axes. (Bottom) Figure rotation pattern speed fits. We note that the simulation with an impact parameter of $0^\circ$ is identical to the simulation with an infall $\phi_{\text{sat}}=45^\circ$ presented in figure \ref{fig:Model_10_phi_host_x_y_plane_infall}.}
    \label{fig:Model_11_phi_host_x_y_plane_infall}
\end{figure}

The models we presented in section \ref{sec:radial_infall} demonstrate that radial mergers may induce figure rotation whose pattern speeds are dependent on the angle of infall with respect to the major axis. In this section we present a second set of models to probe the effect of satellite orbital eccentricity on the resultant figure rotation. We quantify eccentricity by the angular offset between the satellite's initial velocity and radial infall, which we define to be the `impact angle' and measure as
\begin{equation}
    \psi_\mathrm{sat} = \arcsin(-\hat{r}_{\mathrm{sat},0}\times\hat{v}_{\mathrm{sat},0}),
\label{eq:impact_param}
\end{equation}
where $\hat{r}_{\mathrm{sat},0}$ and $\hat{v}_{\mathrm{sat},0}$ are unit vectors pointing in the direction of the satellite's initial position and velocity, respectively. We note that as $\psi$ increases $\hat{v}_0$ swings clockwise in the orbital plane. The models we test have impact angles of 0$^\circ$, 10$^\circ$ , 20$^\circ$, and 30$^\circ$. Each of these models has the same $N-$body Schwarzschild model for the host and a 1:3 merger mass ratio, with the satellite beginning 138 kpc from the host $45^\circ$ from the major axis in the plane containing the major and intermediate axes.

We use $\psi_\mathrm{sat}$ to parametrize these models because it is important to specify the trajectory of the satellite with respect to the major axis; $\psi_\mathrm{sat}>0$ corresponds to a model whose satellite is orbiting counter-clockwise and initially traveling away from the host major axis ($\dot{\phi}_\mathrm{sat}>0$). This orbit will decay differently and therefore torque the host halo differently than one for which $\psi_\mathrm{sat}<0$ with a satellite moving initially clockwise towards the host major axis ($\dot{\phi}_\mathrm{sat}<0$). The initial orbital eccentricity does not capture this important distinction, but we report the corresponding initial eccentricities for our models with $\psi_\mathrm{sat} = 0^\circ,$ $10^\circ$ , $20^\circ$, and $30^\circ$ are $0.99$, $0.97$, $0.93$, and $0.87$. The radial infall model with $\psi_\mathrm{sat} = 0^\circ$ is not perfectly eccentric because the satellite is torqued by the triaxial host.

In the top panel of figure \ref{fig:Model_11_phi_host_x_y_plane_infall} we show the position angle of the host major axis $\varphi$ over the 8 Gyr duration of the simulations. As in our radial infall simulations, the host major axis exhibits a rapid transient 'swinging' response during the earlier phases of the merger, along with a steady tumbling after the merger is concluded. The transient phase is extended in these simulations relative to our radial infall simulations because the higher impact angles allow the satellite to persist for a longer time in the halo before the eventual merger. The pattern speeds post-merger (between $15^\circ-40^\circ$ Gyr$^{-1}$, bottom panel of figure \ref{fig:Model_11_phi_host_x_y_plane_infall}) are higher than those seen from radial mergers ($2^\circ-5^\circ$ Gyr$^{-1}$, see figure \ref{fig:Model_10_phi_host_x_y_plane_infall}). 

Although these systems receive additional angular momentum from the orbit of the satellite compared to the radial infall systems, we suggest that these higher pattern speeds are the product of their longer merger timescales. Because the satellite persists longer in the host halo, it is able to torque the host halo over a proportionally longer timescale. As evidence for this, we call the reader's attention to our model with $\psi_\mathrm{sat}=20^\circ$, which has the highest post-merger pattern speed despite having less orbital angular momentum than our model with $\psi_\mathrm{sat}=30^\circ$. This is a reasonable outcome if the tidal torquing from the satellite is responsible for the post-merger tumbling, because the model with $\psi_\mathrm{sat}=20^\circ$ will have closer pericenter passages and hence higher peak torques than with $\psi_\mathrm{sat}=30^\circ$. The rate at which the halo tumbles in this scenario is evidently a trade-off between the timescale over which the satellite is able to torque the host halo and the maximum strength of the torques (determined in part by the proximity of the satellite's first several pericenters). Taken together, our simulations demonstrate that massive mergers may induce figure rotation that is long-lived in isolation, that tidal torquing is responsible for the resultant figure rotation, and that the orbit of the satellite can influence the pattern speed of rotation long after the merger is concluded.

\subsection{TNG50 MW analogs}

\subsubsection{Dipole and quadrupole power}

\begin{figure*}
    \centering
    \includegraphics[width=0.75\textwidth]{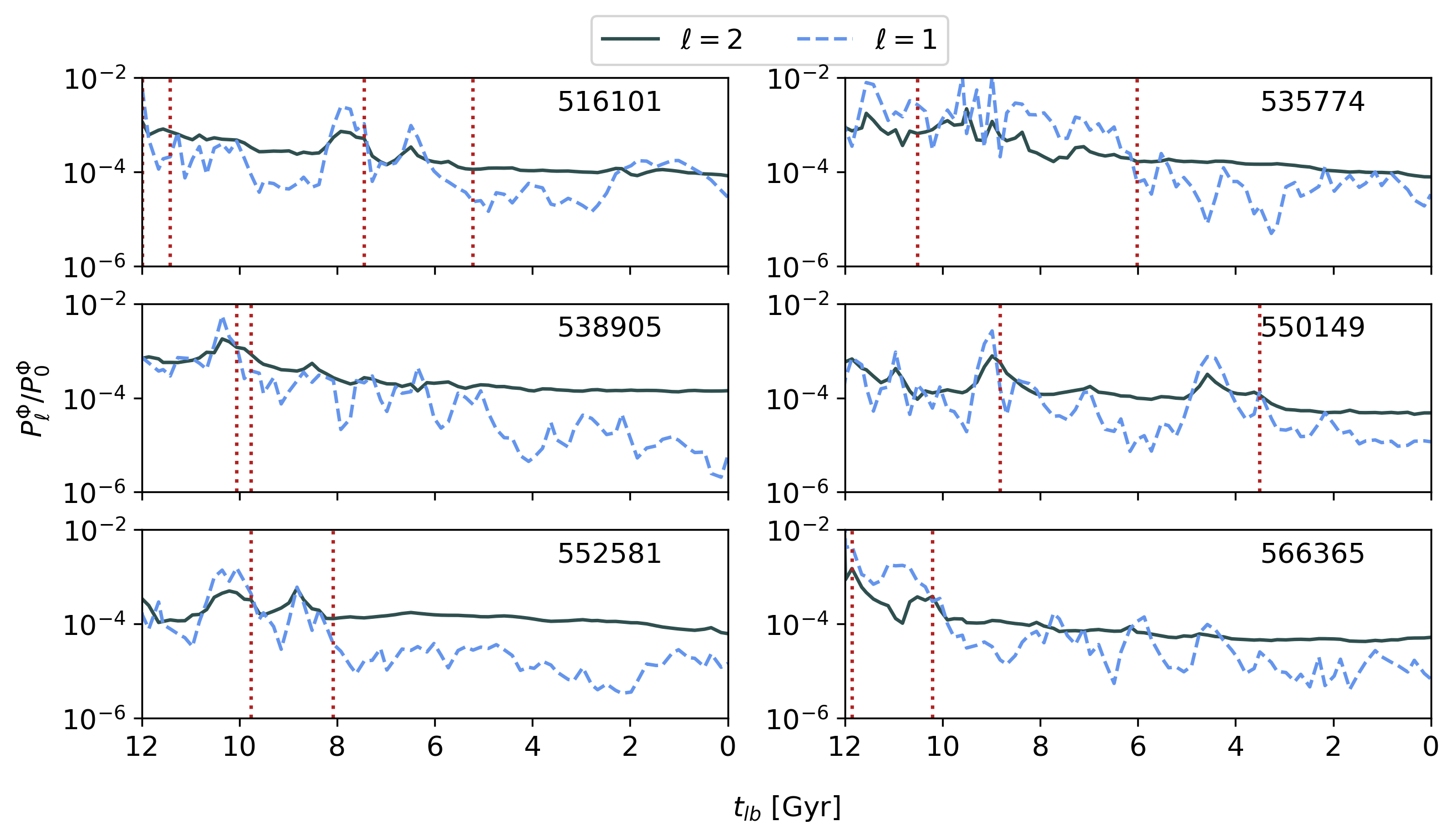}
    \caption{Average power in the quadrupole (solid line) and dipole (dashed line) coefficient series normalized by the monopole term as a function of lookback time. The average power is calculated over radial bins in the same range used for figure rotation fitting. The conclusion of major mergers (i.e. those whose mass ratio $q\geq1:10$) are marked by the vertical dotted lines.}
    \label{fig:MW_best_coeff_power}
\end{figure*}

Among the 198 MW and M31 analogs identified by \cite{pillepich_milky_2024}, we restrict ourselves to the 6 identified as the closest MW analogs and generate multipole expansions for their density distributions and potentials to trace their evolution over the duration of the simulation. In figure \ref{fig:MW_best_coeff_power}, we show the time-evolving averaged power in the dipole (dashed line) and quadrupole (solid line) potential series, normalized by the monopole term. We compute the average over the same radial range as for our pattern speed fitting, beginning where the radial velocity switches from baryon-dominated to DM-dominated and truncating at 60\% of the virial radius, each measured at present day. Alongside the averaged coefficient power, we show the times at which major mergers conclude. We define a major merger as those whose mass ratio $\geq1:10$, and mark them as concluded at the last snapshot at which Subfind is able to identify the subhalo. 

The dipole and quadrupole power series jointly demonstrate the presence or absence of substructure within the halo. Within 2 Gyr prior to any major merger, both the quadrupole and dipole power rise substantially in response to the infalling satellite, often with the dipole power exceeding the quadrupole power in agreement with eq. \ref{eq:subhalo_power_series}. In quiescence, the quadrupole power is either stable or modestly decreasing, in agreement with \cite{arora_shaping_2025}. During these times, the dipole power remains 1-2 dex lower with respect to the quadrupole. Early in the evolution of each of these halos there is a strongly merger-dominated era during which the dipole and quadrupole powers are comparable in magnitude over an extended time. Subhalo 535774 appears to have the noisiest accretion history, with a dipole power consistently or nearly consistently above the quadrupole until its last major merger concludes near $t_{lb}=6$ Gyr. Subhalo 516101 has a relatively quiet assembly history, and appears relatively unperturbed from $10\gtrsim t_{lb}\gtrsim8$ Gyr. Subhalos 516101, 538905, and 535774 appear to become perturbed again several Gyr following their last major merger. This is possibly due to a minor merger or an extended encounter with a massive subhalo which has not merged by present day. While it remains possible using our methodology to fit a rotation to the quadrupole coefficient series during an ongoing merger, the strongest signal of figure rotation will be present when the halo is relatively unperturbed by substructure.

\subsubsection{Quadrupole tumbling}
\begin{figure*}
    \centering
    \includegraphics[width=0.75\textwidth]{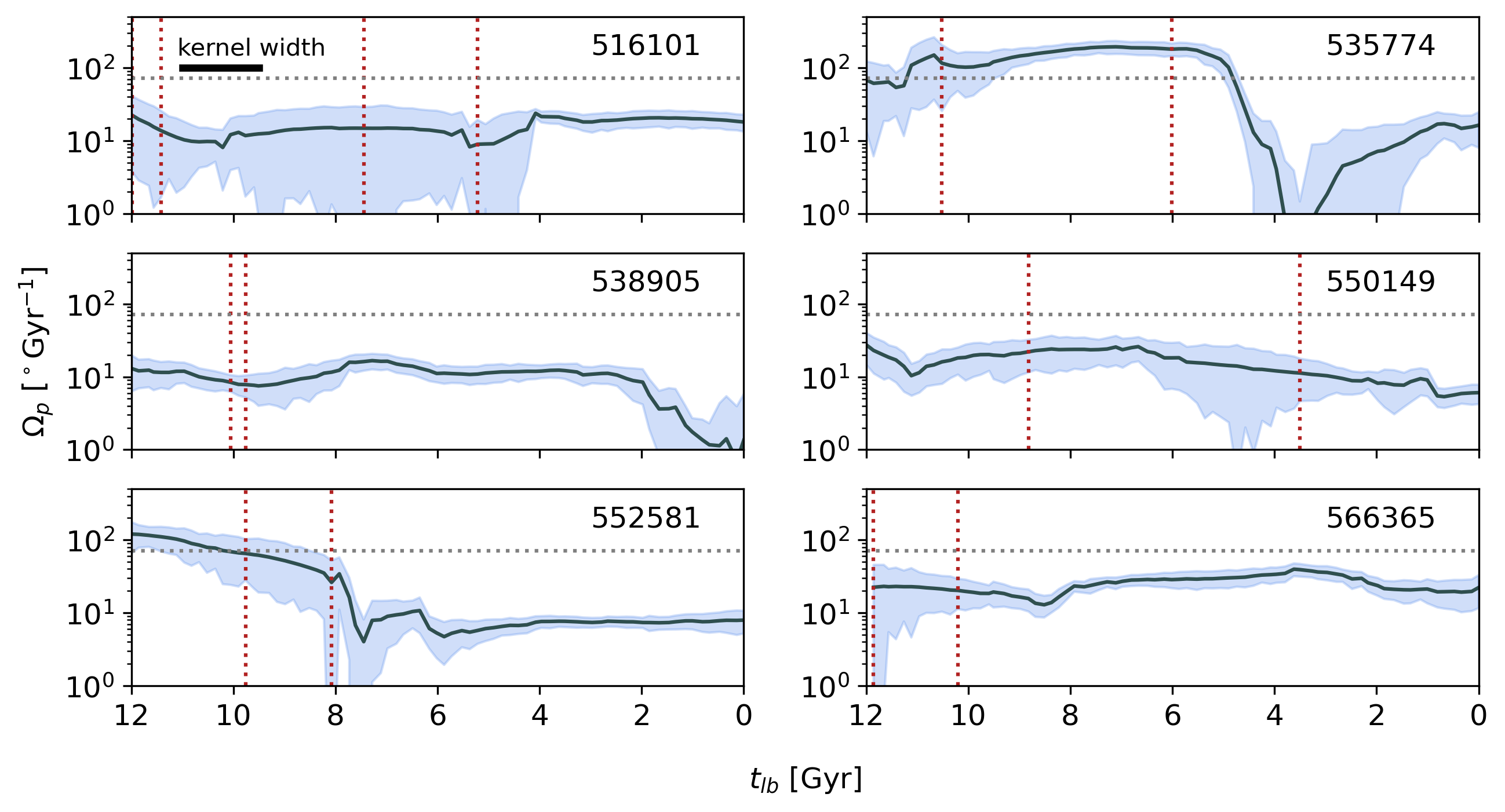}
    \caption{Quadrupole rotation pattern speed fits using the multipole expansion potential coefficients for the six closest MW analogs identified by \cite{pillepich_milky_2024}. Bootstrap fit uncertainties are indicated by the blue shaded region. The vertical red dotted lines mark the conclusion of massive mergers ($\geq1:10$), and the gray horizontal dotted line marks the stability limit for figure rotation determined by \cite{deibel_orbital_2011}.}
    \label{fig:MW_best_analog_PS_fits}
\end{figure*}

\begin{figure*}
    \centering
    \includegraphics[width=.65\textwidth]{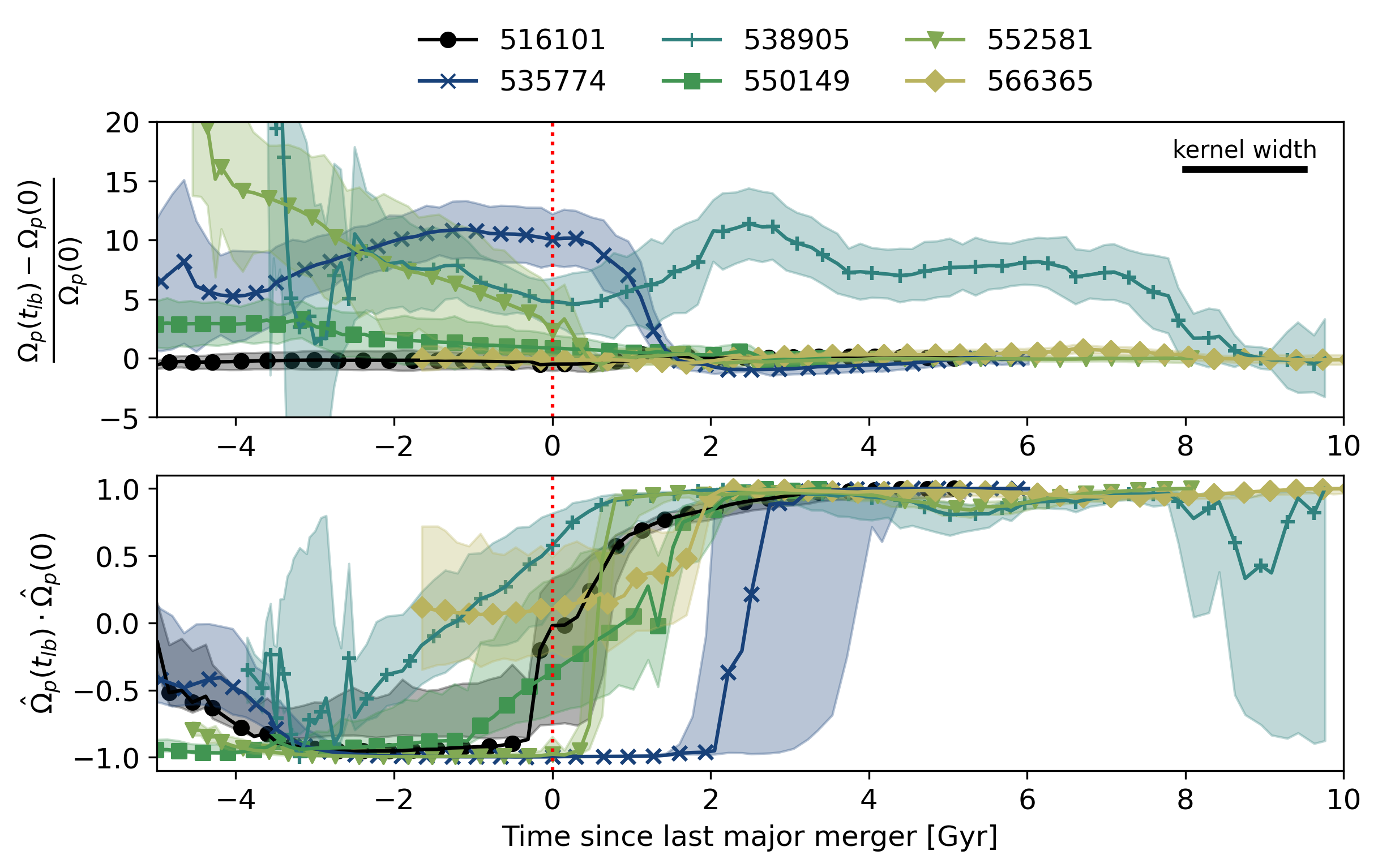}
    \caption{Change in quadrupole rotation pattern speed (top panel) and rotation axis orientation (bottom panel) over time since the last major merger (i.e. whose mass ratios $\geq1:10$) for each of the 6 TNG50 Milky Way analogs identified by \cite{pillepich_milky_2024}, measured with respect to the present-day values. Shaded regions mark the bootstrap fit uncertainties. The standard deviation of the Gaussian kernel used for local pattern speed fits ($\sigma=1.5$ Gyr) is marked for reference in the top left panel. Data markers are shown only at every other snapshot so that the symbols may be distinguished. Evolution in both the pattern speed and rotation axis orientation ceases within $2-4$ Gyr following the last major merger for all but one of our six galaxies.}
    \label{fig:MW_best_time_norm_PS_fits}
\end{figure*}

We apply our quadrupole rotation fitting method discussed in section \ref{sec:methods_rotation_fitting} to each of the six MW analogs and display the resulting pattern speeds in figure \ref{fig:MW_best_analog_PS_fits}. The quadrupole pattern speeds for these six halos show fairly distinct evolution over lookback time, in some cases appearing strongly influenced by the occurrence of major mergers and in other cases showing little evolution around mergers. At present day, the pattern speeds we observe for the quadrupole fall between $0^\circ-33^\circ$ Gyr$^{-1}$ with a median of $12^\circ$ Gyr$^{-1}$ and are consistent with past estimates of figure rotation pattern speeds for isolated $\Lambda$CDM halos \citep{ash_figure_2023,bryan_figure_2007,bailin_figure_2004,dubinski_cosmological_1992}.

The halos of galaxies 535774 and 552581 each experience rapid quadrupole pattern speeds between $100^\circ-200^\circ$ Gyr$^{-1}$ during certain periods of their evolution, well in excess of the figure rotation stability limit of \cite{deibel_orbital_2011}, above which box orbits supporting figure rotation become destabilized and the halo shape begins to evolve. Indeed, halos 535774 and 552581 are clearly in disequilibrium during these times, as marked by the high dipole power (see fig. \ref{fig:MW_best_coeff_power}), suggesting that both halos are merger-dominated during this time. During such times, the quadrupole magnitude and orientation may become dominated by the dynamical friction wake (with some additional contribution from a collective response-like reaction as seen in our idealized simulations, see fig. \ref{fig:aspherical_density_timeseries}), and hence it is likely that the high pattern speeds we observe for the quadrupole measure the orbital frequency of the wake material and satellite. However, \cite{ash_figure_2023} observed in a few isolated halos that figure rotation with these pattern speeds could persist for durations $\lesssim 700$ Myr.

In figure \ref{fig:MW_best_time_norm_PS_fits} we show the time evolution of the quadrupole pattern speed (top panel) and rotation axis orientation (bottom panel) since the last major merger for each analog. Each of these are normalized with respect to their present day values, as either the fractional difference from the present day for the pattern speed or the angular offset of each fit rotation axis from the present day rotation axis. The quadrupole rotation fits are strongly influenced by the last major merger for each of our six galaxies, as demonstrated by the significant evolution within $1-2$ Gyr pre- or post-merger in either the fit rotation axis, or in both the fit rotation axis and pattern speed. Of our six galaxies, three show pattern speeds which are significantly modified within $1-2$ Gyr of the merger (535774, 552581, and 53905) and all six show significant reorientation of their rotation axes, often completely flipping their tumbling direction. Because of the Gaussian kernel used for our rotation fitting, we do not accurately resolve the evolution of quadrupole tumbling on timescales shorter than $\sim1.5$ Gyr. However, we are still able to capture significant evolution in the vicinity of the last major merger for each of our six halos.

Beginning $2-4$ Gyr after the last major merger, the axis and pattern speed of the quadrupoles stabilize and remain effectively constant until $t_{lb}=0$ for every galaxy except 538905. The halo of 538905 instead shows steady tumbling for $\sim4$ Gyr ($\sim4$ Gyr post-merger until $\sim8$ Gyr post-merger, or between $6\gtrsim t_{lb}\gtrsim 2$) following its last major merger before slowing down until $t_{lb}=0$, at which point it evidently ceases to tumble (see the left middle panel of fig. \ref{fig:MW_best_analog_PS_fits}). Consistent with our idealized simulations, the pattern speeds $\sim1$ Gyr pre-merger are frequently $5-10\times$ greater than the post-merger pattern speeds. This is possibly indicative of the transient `swinging' response observed in our idealized simulations. We do not observe any systems in which the pattern speeds pre-merger are significantly slower compared to post-merger. This is likely explained by the series of hierarchical mergers each halo suffers during its early formation ($12\gtrsim t_{lb}\gtrsim8$ Gyr), which will continually torque and reorient the halo quadrupole.

\subsubsection{Subhalo 552581}

\begin{figure}[b]
    \centering
    \includegraphics[width=1\linewidth]{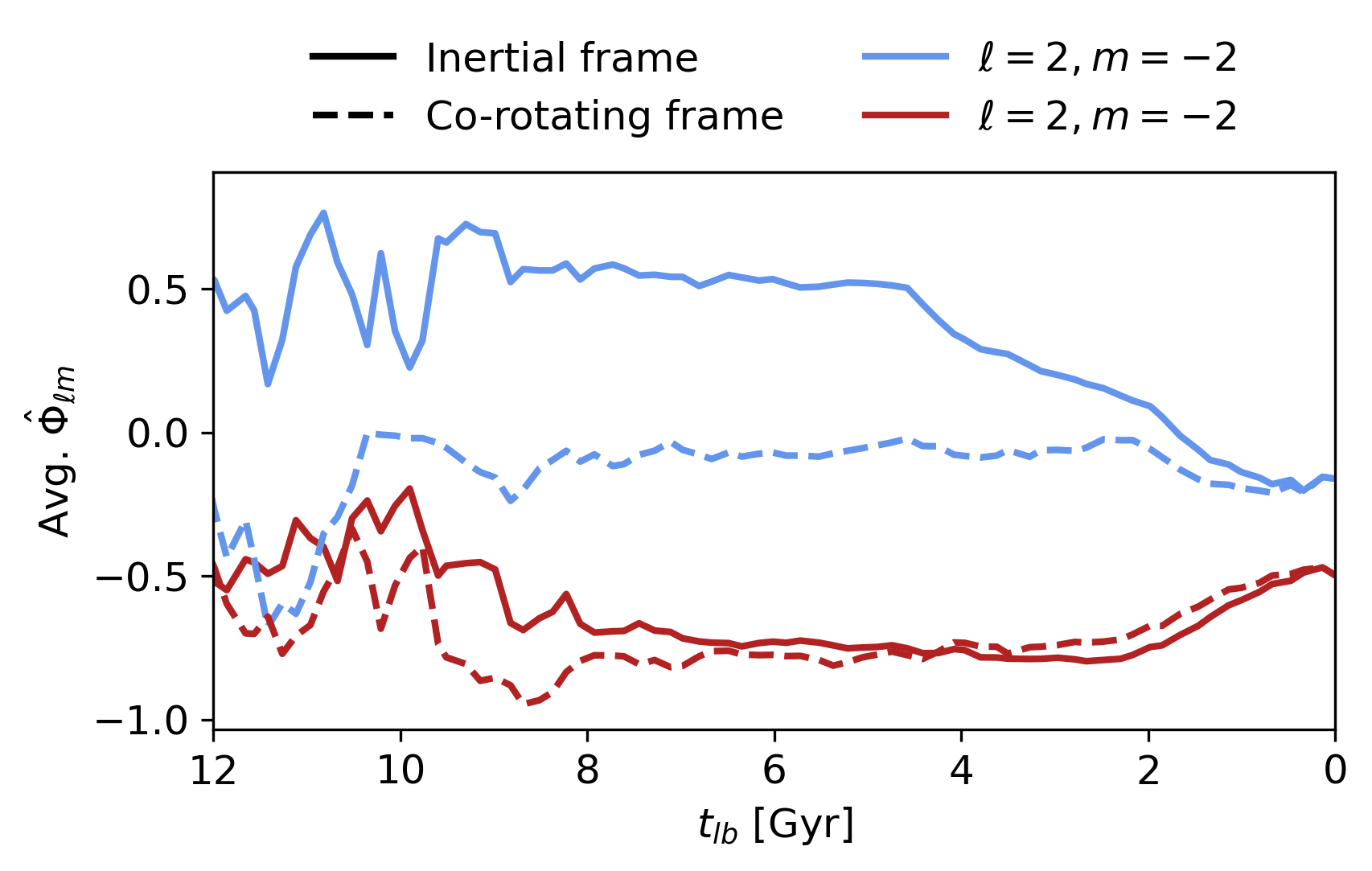}
    \caption{$m=\pm2$ normalized quadrupole potential coefficients for Subhalo 552581, averaged over the radial range used for our quadrupole rotation fitting routine, in both the inertial (solid line) and co-rotating (dashed line) frames, against lookback time. The co-rotating frame was identified by our quadrupole rotation fitting method.}
\label{fig:Subhalo_552581_coefficient_evolution_inertial_vs_corot}
\end{figure}

\begin{figure*}
    \centering
    \includegraphics[width=0.75\textwidth]{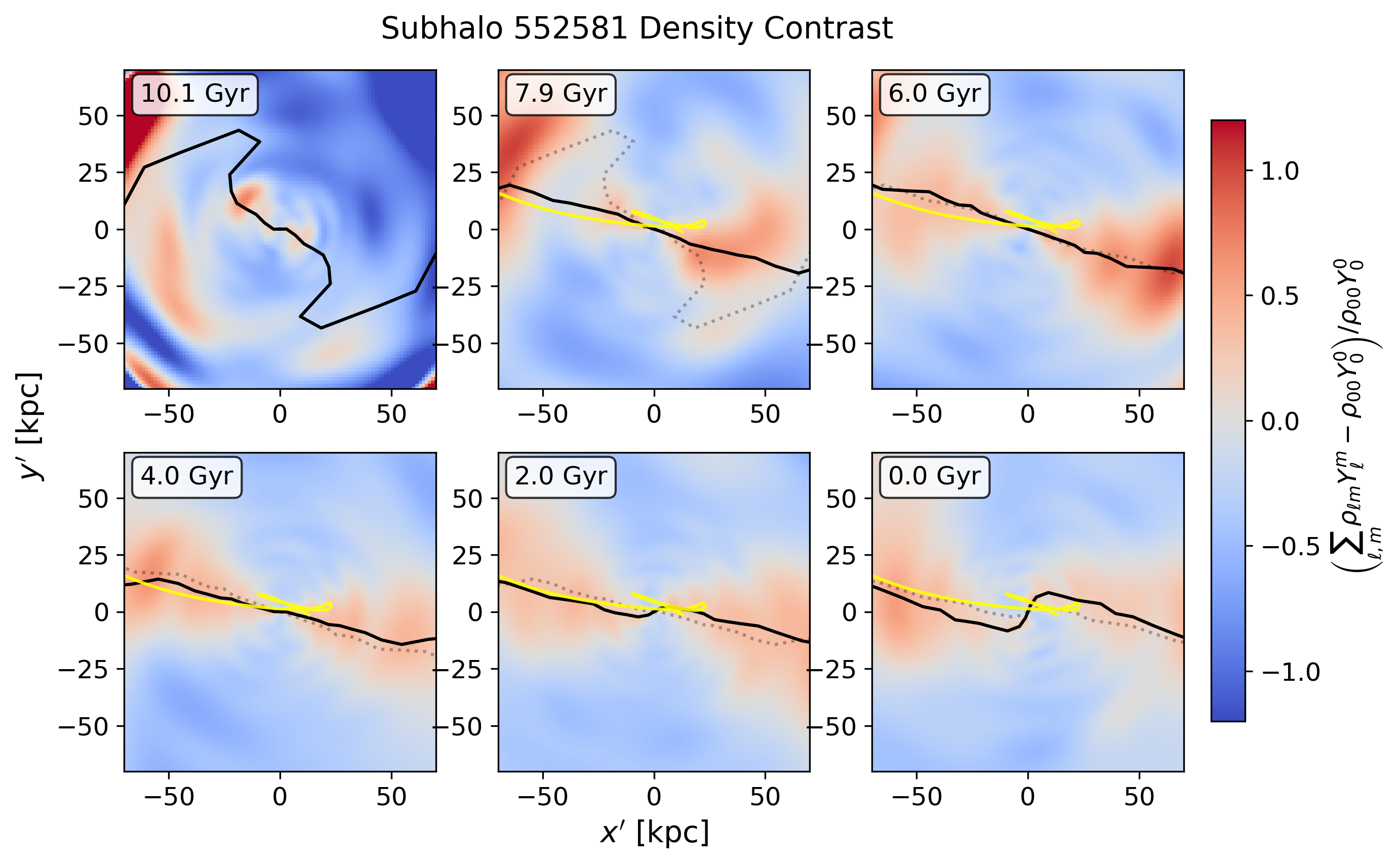}
    \caption{Density contrast plots for Subhalo 552581. The yellow line shows the trajectory of the halo's last major merger (concluded by $t_{lb} = 8$ Gyr). The solid black line shows the plane-projected quadrupole orientation, and the dotted black line shows the same quantity 2 Gyr prior for reference. The merger changes the orientation of the outer halo quadrupole by its conclusion. The halo begins to tumble beginning after $t_{lb}=6$ Gyr, evidently tumbling faster in its inner regions and developing a strong tilt. This tilt manifests as a sharp turn in the plane-projected quadrupole (black line) and begins to develop after $t_{lb}=6$ Gyr at small galactocentric radii, expanding outwards to $\sim20$ kpc and nearly $90^\circ$ at $t_{lb}=0$.}
    \label{fig:Subhalo_552581_density_contrast}
\end{figure*}

As a case study, we present more detailed results for Subhalo 552581. In figure \ref{fig:Subhalo_552581_coefficient_evolution_inertial_vs_corot} we show the normalized quadrupole potential coefficients $m=\pm2$ for Subhalo 552581 averaged over the radial range used for rotation fitting against lookback time. We plot these averaged coefficients both in the inertial and co-rotating frames. The coefficients show relatively little evolution in the co-rotating frame from $8 \geq t_{lb} \geq 2$~Gyr in both coefficient series, indicating that much of the time evolution in the inertial frame is due to figure rotation and that our best-fit quadrupole tumbling model does a good job capturing the coefficient time dependence over this time interval. The observed evolution for $t_{lb}\geq8$ Gyr in the co-rotating frame is associated with the merger, and is an indication that the quadrupole time evolution in the vicinity of the merger is not well-described by smooth figure rotation varying on time-scales of $\sim1.5$ Gyr (i.e., the kernel width used in our fitting). This behavior is expected; we can observe in figure \ref{fig:MW_best_coeff_power} that the changes to the amplitude of the halo quadrupole associated with the merger around this time proceed on timescales much shorter than $1.5$ Gyr, as did the orientation of the quadrupole during ongoing mergers in our idealized simulations (fig. \ref{fig:aspherical_density_timeseries}).

To demonstrate in more detail the evolution of the halo 552581's density structure, we plot its mid-plane density contrast (defined as the fractional deviation from the spherically-averaged density) in figure \ref{fig:Subhalo_552581_density_contrast} in the $x-y$ plane of the same reference frame in which we performed our multipole expansion. As before, we plot the plane-projected quadrupole orientation (black line) at the current time as well as 2 Gyr prior (dotted line). At $t_{lb}=10$ Gyr (top left panel), the halo is merger-dominated and density contrast shows a predominantly dipolar shape, albeit with varying orientations, in the outer halo. Once the merger has concluded at $t_{lb}=8$ Gyr (top middle panel), the density contrast takes on a more quadrupolar shape which is evidently rotated into alignment with the merger's infall trajectory. The quadrupole is strengthened by $t_{lb}=6$ Gyr, after which its orientation gradually tumbles within the plane by a few degrees every 2 Gyr\footnote{We have not aligned these frames to the plane in which this halo tumbles, and hence the projected tumbling rate will be slower than or equal to our measured tumbling rate.}. Matching our observation in the coefficient series, the inner halo appears to tumble somewhat faster than the outer halo during this time. As the inner halo does so it torques the surrounding halo, speeding up the rotation in the surrounding region and causing the halo to develop a pronounced tilt around 20 kpc. It is likely that the disk and inner DM halo of Subhalo 552581 are tilting coherently during this period, possibly as a result of torques from the merger debris \citep{dodge_dynamics_2023}. The presence of a steadily tumbling quadrupole with a consistent phase angle and pattern speed between $10-70$ kpc over the last 6 Gyr provides confirmation that the quadrupole tumbling in this halo traces a rotation in the entire halo shape, and hence represents halo figure rotation.

\section{Discussion and Conclusion}\label{sec:discussion}

We have investigated the role of massive mergers in inducing dark halo figure rotation in both idealized, isolated $N-$body simulations and in MW analogs from TNG50 of the IllustrisTNG simulation suite. Our key findings are as follows:
\begin{enumerate}
    \item Massive mergers are capable of inducing dark halo figure rotation in isolation. Figure rotation can be induced even for purely radial mergers which contribute no angular momentum to the host halo, and the pattern speed of rotation is dependent on the orbit of the infalling satellite.

    \item Merger-induced halo tumbling occurs in two phases: a rapid transient phase during the satellite's infall, and a long-lived phase after the mergers conclusion. The early transient phase has been previously observed \citep[e.g.,][] {arora_shaping_2025} and is characterized by pattern speeds which are $5-10\times$ higher than the long-lived pattern speeds. The long-lived rotation may remain stable for a Hubble time or longer post-merger and shows pattern speeds consistent with those observed in $\Lambda$CDM cosmological simulations.
    
    \item The re-orientation of the halo quadrupole during a merger proceeds outwards-in in our idealized simulations. During in-fall, the quadrupole orientation is strongly influenced by the dynamical friction wake. In the final stages of the merger the quadrupole experiences a strong torque from the satellite and becomes aligned to the satellite debris in the inner halo ($\lesssim 15$ kpc in our simulations). This change in the quadrupole orientation propagates outwards through the halo as the inner regions torque the outer halo, and appears capable of inducing twists in the host halo which endure for at least a Gyr.
    
    \item Satellites infalling on more tangential orbits induce higher pattern speeds in some of our models. This is likely because orbital angular momentum extends the timescale for the satellite's orbit to decay, allowing it to torque the host halo for a longer time. The highest tumbling rates we observed in our idealized simulations occurred at intermediate impact angles, which appeared to create an optimal balance between the maximum tidal torque (set by the radius of the first 1-2 pericenters) and the satellite orbital decay timescale.    
    
    \item Major mergers strongly influence both the quadrupole pattern speed in three of our six halos and rotation axis orientation in all six halos of our TNG50 MW analog galaxies identified by \cite{pillepich_milky_2024}. The quadrupole pattern speeds and rotation axes stabilize to their present day values in all but one of the six MW analog galaxies within  $2-4$ Gyr after the last major merger has concluded and remain stable until present day. The one galaxy which is an exception to this shows stable rotation over a 4 Gyr period following its last major merger. 
\end{enumerate}

This paper represents the first attempt, to our knowledge, to investigate in detail whether major mergers are capable of significantly affecting the steady tumbling of $\Lambda$CDM halos, or `figure rotation'. Figure rotation has been robustly predicted for decades by studies of $\Lambda$CDM cosmological simulations \citep{dubinski_cosmological_1992,bailin_figure_2004,bryan_figure_2007,ash_figure_2023}. Prior to this work, tidal shear from large scale structure during initial halo collapse and virialization was the only mechanism known to produce figure rotation \citep{dubinski_cosmological_1992}, \textedit{though \cite{drakos_major_2019} note tumbling in the major axis orientation following a merger in at least one of their idealized simulations}. Our findings therefore represent an important paradigm shift; we have demonstrated using idealized simulations that, besides tidal shear during halo collapse, massive mergers are additionally capable of inducing figure rotation in isolation in a halo that is initially non-tumbling. Furthermore, in the vicinity of the last major merger, three of our six TNG50 cosmological halos show quadrupole pattern speeds that are significantly altered, and all six show substantial evolution of their quadrupole tumbling axis. This suggests that, even in a cosmological setting where rotation is initialized during halo collapse, major mergers may \textit{reset} the axis and pattern speed of figure rotation. Present day figure rotation may therefore instead be representative of the last major merger rather than primordial tidal shear. However, evolution of the tumbling axis and pattern speed after the last major merger are still possible, as seen in some of our idealized models (fig. \ref{fig:Model_11_phi_host_x_y_plane_infall}) and one of our cosmological halos (see subhalo 538905 in figs. \ref{fig:MW_best_analog_PS_fits} and \ref{fig:MW_best_time_norm_PS_fits}). \cite{ash_figure_2023} also observed significant evolution of figure rotation in many isolated halos without significant substructure. 


\textedit{It has been previously noted that a robust detection figure rotation may useful in distinguishing between particle-based DM and modified gravity models such as MOND \citep{bailin_figure_2004,valluri_detecting_2021,ash_figure_2023}, making it an intriguing and relatively unexplored cosmological probe. In this work we establish that figure rotation may be connected to a galaxy's merger history, implying that it may also be sensitive to cosmological parameters affecting halo assembly history such as the matter power spectrum normalization $\sigma_8$ \citep{amoura_halo_2023}. While, to the best of our knowledge, figure rotation has only been studied in $\Lambda$CDM simulations run in a narrow grid of cosmological parameters, future work could probe a wider parameter grid of $\Lambda$CDM simulations or alternative DM models.}

It has been previously shown that major mergers are capable of causing stellar disks to either tumble or turn over \citep[][]{dillamore_merger-induced_2022,dodge_dynamics_2023,bell_galaxy_2026} with steady tilting rates between $5^\circ-10^\circ$ Gyr$^{-1}$ or rapidly reorienting (``flipping'') with tilting rates as high as $60^\circ$ Gyr$^{-1}$ or up to $200^\circ$ Gyr$^{-1}$. These disk tilting rates closely match the expected pattern speeds for halo figure rotation, for which steady tumbling rates $\sim10^\circ$ Gyr$^{-1}$\citep{dubinski_cosmological_1992, bailin_figure_2004, bryan_figure_2007, ash_figure_2023} and short-lived rapid reorientation $\sim60^\circ-300^\circ$ Gyr$^{-1}$ \citep{ash_figure_2023} have each been observed in cosmological simulations. Both the steady tumbling rates and short-lived rapid reorientation rates are in good agreement with the quadrupole tumbling rates we observe in our idealized simulations (figs. \ref{fig:Model_10_phi_host_x_y_plane_infall} and \ref{fig:Model_11_phi_host_x_y_plane_infall}) and in the TNG50 MW analogs (fig. \ref{fig:MW_best_analog_PS_fits}), suggesting that the disk tilting and halo figure rotation may indeed be closely linked. This is almost certainly the case in the vicinity of the disk, where the shape and orientation of the DM are coupled to the disk via adiabatic contraction \citep[e.g.,][]{kazantzidis_effect_2004, gnedin_response_2004, chua_shape_2019,prada_dark_2019} as well as resonant trapping with the bar \citep[e.g.,][]{collier_coupling_2021,marostica_response_2024} and spiral arms \citep[e.g.,][]{bernet_dark_2025}. \cite{dillamore_merger-induced_2022} found that, in addition to the inner DM halo minor axis ($<30$ kpc) remaining closely aligned to the stellar disk axis as the disk re-aligned, the outer halo minor axis evolved semi-coherently with the disk and inner halo, remaining aligned to the inner halo to within $\sim25^\circ$ over roughly 10 Gyr of evolution post-merger. The ability of mergers to induce disk tilting taken together with our findings in this work suggest that, rather than been independent phenomena, disk tilting and halo figure rotation may commonly be coupled motions. 

We have deliberately restricted ourselves to investigating major mergers which have concluded by present day, and are therefore predominantly ancient. However, our own Galaxy is experiencing an ongoing massive interaction with the Large Magellanic Cloud (LMC) with an expected mass ratio between $1:10$ and $1:5$ (see \cite{vasiliev_effect_2023} for a review). The LMC is likely significantly perturbing the MW halo by inducing both an overdense wake tracing the LMC's past orbit and a dipolar density perturbation in the stellar halo \citep{garavito-camargo_quantifying_2021}. While we do not investigate low-redshift massive interactions like that between the MW and LMC, our idealized simulations suggest that the LMC may be capable of inducing a transient `swinging' response in the MW's halo, even if it has only undergone 1-2 pericentric passages. Such a response is also observed by \cite{arora_shaping_2025}. This response could represent an additional perturbation on the MW's halo exerted by the LMC. In a companion paper, we investigate massive LMC-like encounters using the MW+LMC analog galaxies identified by \cite{pillepich_milky_2024} in TNG50.

In this work we have focused on massive mergers whose mass ratio is $\geq1:10$. Such mergers have been found to increase in prevalence with redshift up to $z=6$ \citep{conselice_structures_2008,lotz_major_2011,bluck_structures_2012,ownsworth_minor_2014,lopez-sanjuan_alhambra_2015,man_resolving_2016,mundy_consistent_2017,ventou_muse_2017,mantha_major_2018,duncan_observational_2019,conselice_direct_2022,patton_interacting_2024,duan_galaxy_2025} and hence are predominantly representative of ancient mergers like those we mark in the TNG50 MW analogs. In the Milky Way, the last major merger occurred at $z\sim2$ with the Gaia-Sausage-Enceladus (GSE) galaxy \citep{belokurov_co-formation_2018,helmi_merger_2018} the debris of which makes up a majority of the inner stellar halo \citep{naidu_evidence_2020}, roughly $20\%$ of the MW's DM, and suggests an initially retrograde infall which decayed to deposit debris on radial orbits through the inner halo \citep{naidu_reconstructing_2021}. Both our idealized simulations and the TNG50 MW analogs suggest that mergers like GSE can induce or alter tumbling in the DM halo, and it is perhaps likely that figure rotation in the MW's halo, if present, is the result of this ancient merger. GSE has previously been suggested as a cause of possible tilting in the MW's disk at rates which may be measurable with \textit{Gaia} astrometry \citep{dillamore_merger-induced_2022,dodge_dynamics_2023}. Furthermore, both our idealized simulations and cosmological halos demonstrate that it is possible for figure rotation to remain stable for at least a Hubble time following the last major merger, suggesting that potential measurements of either figure rotation or disk tilting in the Milky Way could provide a novel dynamical probe of the early MW-GSE merger which occurred $8-10$ Gyr ago.

\begin{acknowledgements}
We thank members of the Stellar Halos group at the University of Michigan and Eugene Vasiliev for helpful suggestions and discussion through out the course of this work. We thank the IllustrisTNG collaboration for access to the TNG50 simulation suite and compute time via the JupyterLab environment hosted at the Max Planck Computing and Data Facility.  We gratefully acknowledge support  from NASA ATP grant 80NSSC24K0938 and NSF grant 2510879.
\end{acknowledgements}

\software{AGAMA \citep{vasiliev_agama_2019},
NumPy \citep{harris_array_2020}, 
SciPy \citep{virtanen_scipy_2020}, 
Astropy \citep{the_astropy_collaboration_astropy_2018}, 
matplotlib \citep{hunter_matplotlib_2007}, spherical \citep{boyle_spherical_2023}, quaternionic \citep{boyle_quaternionic_2024}. We used the Perplexity\footnote{https://www.perplexity.ai/} LLM while revising this manuscript to generate mock referee reports, some of the suggestions from which were implemented into the text by authors. All of the writing and analysis presented in this manuscript was performed by the authors and not by any LLM.}

\bibliography{references}{}

@article{amoura_halo_2023,
	title = {Halo growth and merger rates as a cosmological test},
	volume = {527},
	copyright = {https://creativecommons.org/licenses/by/4.0/},
	issn = {0035-8711, 1365-2966},
	url = {https://academic.oup.com/mnras/article/527/2/3459/7371671},
	doi = {10.1093/mnras/stad3416},
	language = {en},
	number = {2},
	urldate = {2026-08-19},
	journal = {Monthly Notices of the Royal Astronomical Society},
	author = {Amoura, Yuba and Drakos, Nicole E and Berrouet, Anael and Taylor, James E},
	month = nov,
	year = {2023},
	pages = {3459--3473},
}

@article{power_inner_2003,
	title = {The inner structure of {CDM} haloes -- {I}. {A} numerical convergence study},
	volume = {338},
	issn = {0035-8711, 1365-2966},
	url = {https://academic.oup.com/mnras/article/338/1/14/1094443},
	doi = {10.1046/j.1365-8711.2003.05925.x},
	language = {en},
	number = {1},
	urldate = {2026-08-19},
	journal = {Monthly Notices of the Royal Astronomical Society},
	author = {Power, C. and Navarro, J. F. and Jenkins, A. and Frenk, C. S. and White, S. D. M. and Springel, V. and Stadel, J. and Quinn, T.},
	month = jan,
	year = {2003},
	pages = {14--34},
}

@article{zhan_optimal_2006,
	title = {Optimal {Softening} for \textit{{N}} ‐{Body} {Halo} {Simulations}},
	volume = {639},
	issn = {0004-637X, 1538-4357},
	url = {https://iopscience.iop.org/article/10.1086/499763},
	doi = {10.1086/499763},
	language = {en},
	number = {2},
	urldate = {2026-08-19},
	journal = {The Astrophysical Journal},
	author = {Zhan, Hu},
	month = mar,
	year = {2006},
	pages = {617--620},
}

@article{mcmillan_haloes_2007,
	title = {The haloes of merger remnants},
	volume = {376},
	issn = {0035-8711, 1365-2966},
	url = {https://academic.oup.com/mnras/article-lookup/doi/10.1111/j.1365-2966.2007.11516.x},
	doi = {10.1111/j.1365-2966.2007.11516.x},
	language = {en},
	number = {3},
	urldate = {2026-08-19},
	journal = {Monthly Notices of the Royal Astronomical Society},
	author = {McMillan, P. J. and Athanassoula, E. and Dehnen, W.},
	month = apr,
	year = {2007},
	pages = {1261--1269},
}

@article{drakos_major_2019,
	title = {Major mergers between dark matter haloes – {I}. {Predictions} for size, shape, and spin},
	volume = {487},
	copyright = {https://academic.oup.com/journals/pages/open\_access/funder\_policies/chorus/standard\_publication\_model},
	issn = {0035-8711, 1365-2966},
	url = {https://academic.oup.com/mnras/article/487/1/993/5488452},
	doi = {10.1093/mnras/stz1306},
	language = {en},
	number = {1},
	urldate = {2026-08-12},
	journal = {Monthly Notices of the Royal Astronomical Society},
	author = {Drakos, Nicole E and Taylor, James E and Berrouet, Anael and Robotham, Aaron S G and Power, Chris},
	month = jul,
	year = {2019},
	pages = {993--1007},
}

@article{hunter_matplotlib_2007,
	title = {Matplotlib: {A} {2D} {Graphics} {Environment}},
	volume = {9},
	copyright = {https://ieeexplore.ieee.org/Xplorehelp/downloads/license-information/IEEE.html},
	issn = {1521-9615},
	shorttitle = {Matplotlib},
	url = {http://ieeexplore.ieee.org/document/4160265/},
	doi = {10.1109/MCSE.2007.55},
	number = {3},
	urldate = {2026-05-13},
	journal = {Computing in Science \& Engineering},
	author = {Hunter, John D.},
	year = {2007},
	pages = {90--95},
}

@article{schwarzschild_triaxial_1982,
	title = {Triaxial equilibrium models for elliptical galaxies with slow figure rotation},
	volume = {263},
	issn = {0004-637X, 1538-4357},
	url = {http://adsabs.harvard.edu/doi/10.1086/160531},
	doi = {10.1086/160531},
	language = {en},
	urldate = {2026-05-13},
	journal = {The Astrophysical Journal},
	author = {Schwarzschild, M.},
	month = dec,
	year = {1982},
	pages = {599},
}

@article{bell_galaxy_2026,
	title = {Galaxy mergers and disk angular momentum evolution: stellar halos as a critical test},
	volume = {9},
	copyright = {http://creativecommons.org/licenses/by/4.0},
	issn = {2565-6120},
	shorttitle = {Galaxy mergers and disk angular momentum evolution},
	url = {https://astro.theoj.org/article/161450-galaxy-mergers-and-disk-angular-momentum-evolution-stellar-halos-as-a-critical-test},
	doi = {10.33232/001c.161450},
	language = {en},
	urldate = {2026-05-07},
	journal = {The Open Journal of Astrophysics},
	author = {Bell, Eric F. and D'Souza, Richard and Valluri, Monica and Gozman, Katya},
	month = apr,
	year = {2026},
}

@article{eddington_distribution_1916,
	title = {The {Distribution} of {Stars} in {Globular} {Clusters}},
	volume = {76},
	issn = {0035-8711, 1365-2966},
	url = {https://academic.oup.com/mnras/article-lookup/doi/10.1093/mnras/76.7.572},
	doi = {10.1093/mnras/76.7.572},
	language = {en},
	number = {7},
	urldate = {2026-04-23},
	journal = {Monthly Notices of the Royal Astronomical Society},
	author = {Eddington, A. S.},
	month = may,
	year = {1916},
	pages = {572--585},
}

@article{helmi_merger_2018,
	title = {The merger that led to the formation of the {Milky} {Way}’s inner stellar halo and thick disk},
	volume = {563},
	issn = {0028-0836, 1476-4687},
	url = {https://www.nature.com/articles/s41586-018-0625-x},
	doi = {10.1038/s41586-018-0625-x},
	language = {en},
	number = {7729},
	urldate = {2026-04-23},
	journal = {Nature},
	author = {Helmi, Amina and Babusiaux, Carine and Koppelman, Helmer H. and Massari, Davide and Veljanoski, Jovan and Brown, Anthony G. A.},
	month = nov,
	year = {2018},
	pages = {85--88},
}

@article{dodge_dynamics_2023,
	title = {Dynamics of stellar disc tilting from satellite mergers},
	volume = {518},
	copyright = {https://academic.oup.com/journals/pages/open\_access/funder\_policies/chorus/standard\_publication\_model},
	issn = {0035-8711, 1365-2966},
	url = {https://academic.oup.com/mnras/article/518/2/2870/6825513},
	doi = {10.1093/mnras/stac3249},
	language = {en},
	number = {2},
	urldate = {2026-03-02},
	journal = {Monthly Notices of the Royal Astronomical Society},
	author = {Dodge, Benjamin C and Slone, Oren and Lisanti, Mariangela and Cohen, Timothy},
	month = jan,
	year = {2023},
	pages = {2870--2884},
}

@article{nibauer_slant_2024,
	title = {Slant, {Fan}, and {Narrow}: {The} {Response} of {Stellar} {Streams} to a {Tilting} {Galactic} {Disk}},
	volume = {969},
	issn = {0004-637X, 1538-4357},
	shorttitle = {Slant, {Fan}, and {Narrow}},
	url = {https://iopscience.iop.org/article/10.3847/1538-4357/ad4299},
	doi = {10.3847/1538-4357/ad4299},
	number = {1},
	urldate = {2026-03-05},
	journal = {The Astrophysical Journal},
	author = {Nibauer, Jacob and Bonaca, Ana and Lisanti, Mariangela and Erkal, Denis and Hastings, Zoe},
	month = jul,
	year = {2024},
	pages = {55},
}

@misc{johri_disentangling_2026,
	title = {Disentangling {Drivers} of {Disk} {Warps} in {Tilted} and {Tumbling} {TNG50} {Halos}},
	copyright = {Creative Commons Attribution 4.0 International},
	url = {https://arxiv.org/abs/2601.17599},
	doi = {10.48550/ARXIV.2601.17599},
	urldate = {2026-03-05},
	publisher = {arXiv},
	author = {Johri, Saarthak and Ash, Neil and Valluri, Monica},
	year = {2026},
	note = {Version Number: 2},
}

@article{dubinski_warps_2009,
	title = {{WARPS} {AND} {BARS} {FROM} {THE} {EXTERNAL} {TIDAL} {TORQUES} {OF} {TUMBLING} {DARK} {HALOS}},
	volume = {703},
	issn = {0004-637X, 1538-4357},
	url = {https://iopscience.iop.org/article/10.1088/0004-637X/703/2/2068},
	doi = {10.1088/0004-637X/703/2/2068},
	number = {2},
	urldate = {2026-03-05},
	journal = {The Astrophysical Journal},
	author = {Dubinski, John and Chakrabarty, Dalia},
	month = oct,
	year = {2009},
	pages = {2068--2081},
}

@article{vasiliev_effect_2023,
	title = {The {Effect} of the {LMC} on the {Milky} {Way} {System}},
	volume = {11},
	issn = {2075-4434},
	url = {https://www.mdpi.com/2075-4434/11/2/59},
	doi = {10.3390/galaxies11020059},
	language = {en},
	number = {2},
	urldate = {2026-03-04},
	journal = {Galaxies},
	author = {Vasiliev, Eugene},
	month = apr,
	year = {2023},
	pages = {59},
}

@article{naidu_reconstructing_2021,
	title = {Reconstructing the {Last} {Major} {Merger} of the {Milky} {Way} with the {H3} {Survey}},
	volume = {923},
	issn = {0004-637X, 1538-4357},
	url = {https://iopscience.iop.org/article/10.3847/1538-4357/ac2d2d},
	doi = {10.3847/1538-4357/ac2d2d},
	number = {1},
	urldate = {2026-03-04},
	journal = {The Astrophysical Journal},
	author = {Naidu, Rohan P. and Conroy, Charlie and Bonaca, Ana and Zaritsky, Dennis and Weinberger, Rainer and Ting 丁, Yuan-Sen 源森 and Caldwell, Nelson and Tacchella, Sandro and Han, Jiwon Jesse and Speagle, Joshua S. and Cargile, Phillip A.},
	month = dec,
	year = {2021},
	pages = {92},
}

@article{naidu_evidence_2020,
	title = {Evidence from the {H3} {Survey} {That} the {Stellar} {Halo} {Is} {Entirely} {Comprised} of {Substructure}},
	volume = {901},
	issn = {0004-637X, 1538-4357},
	url = {https://iopscience.iop.org/article/10.3847/1538-4357/abaef4},
	doi = {10.3847/1538-4357/abaef4},
	number = {1},
	urldate = {2026-03-04},
	journal = {The Astrophysical Journal},
	author = {Naidu, Rohan P. and Conroy, Charlie and Bonaca, Ana and Johnson, Benjamin D. and Ting 丁, Yuan-Sen 源森 and Caldwell, Nelson and Zaritsky, Dennis and Cargile, Phillip A.},
	month = sep,
	year = {2020},
	pages = {48},
}

@article{belokurov_co-formation_2018,
	title = {Co-formation of the disc and the stellar halo★},
	volume = {478},
	copyright = {http://academic.oup.com/journals/pages/about\_us/legal/notices},
	issn = {0035-8711, 1365-2966},
	url = {https://academic.oup.com/mnras/article/478/1/611/5032586},
	doi = {10.1093/mnras/sty982},
	language = {en},
	number = {1},
	urldate = {2026-03-04},
	journal = {Monthly Notices of the Royal Astronomical Society},
	author = {Belokurov, V and Erkal, D and Evans, N W and Koposov, S E and Deason, A J},
	month = jul,
	year = {2018},
	pages = {611--619},
}

@article{duncan_observational_2019,
	title = {Observational {Constraints} on the {Merger} {History} of {Galaxies} since z ≈ 6: {Probabilistic} {Galaxy} {Pair} {Counts} in the {CANDELS} {Fields}},
	volume = {876},
	issn = {0004-637X, 1538-4357},
	shorttitle = {Observational {Constraints} on the {Merger} {History} of {Galaxies} since z ≈ 6},
	url = {https://iopscience.iop.org/article/10.3847/1538-4357/ab148a},
	doi = {10.3847/1538-4357/ab148a},
	number = {2},
	urldate = {2026-03-03},
	journal = {The Astrophysical Journal},
	author = {Duncan, Kenneth and Conselice, Christopher J. and Mundy, Carl and Bell, Eric and Donley, Jennifer and Galametz, Audrey and Guo, Yicheng and Grogin, Norman A. and Hathi, Nimish and Kartaltepe, Jeyhan and Kocevski, Dale and Koekemoer, Anton M. and Pérez-González, Pablo G. and Mantha, Kameswara B. and Snyder, Gregory F. and Stefanon, Mauro},
	month = may,
	year = {2019},
	pages = {110},
}

@article{ventou_muse_2017,
	title = {The {MUSE} \textit{{Hubble}} {Ultra} {Deep} {Field} {Survey}: {IX}. {Evolution} of galaxy merger fraction since \textit{z} ≈ 6},
	volume = {608},
	copyright = {https://www.edpsciences.org/en/authors/copyright-and-licensing},
	issn = {0004-6361, 1432-0746},
	shorttitle = {The {MUSE} \textit{{Hubble}} {Ultra} {Deep} {Field} {Survey}},
	url = {http://www.aanda.org/10.1051/0004-6361/201731586},
	doi = {10.1051/0004-6361/201731586},
	urldate = {2026-03-03},
	journal = {Astronomy \& Astrophysics},
	author = {Ventou, E. and Contini, T. and Bouché, N. and Epinat, B. and Brinchmann, J. and Bacon, R. and Inami, H. and Lam, D. and Drake, A. and Garel, T. and Michel-Dansac, L. and Pello, R. and Steinmetz, M. and Weilbacher, P. M. and Wisotzki, L. and Carollo, M.},
	month = dec,
	year = {2017},
	pages = {A9},
}

@article{duan_galaxy_2025,
	title = {Galaxy mergers in the epoch of reionization – {I}. {A} {JWST} study of pair fractions, merger rates, and stellar mass accretion rates at \textit{z} = 4.5–11.5},
	volume = {540},
	copyright = {https://creativecommons.org/licenses/by/4.0/},
	issn = {0035-8711, 1365-2966},
	url = {https://academic.oup.com/mnras/article/540/1/774/8115792},
	doi = {10.1093/mnras/staf638},
	language = {en},
	number = {1},
	urldate = {2026-03-03},
	journal = {Monthly Notices of the Royal Astronomical Society},
	author = {Duan, Qiao and Conselice, Christopher J and Li, Qiong and Austin, Duncan and Harvey, Thomas and Adams, Nathan J and Duncan, Kenneth J and Trussler, James and Ferreira, Leonardo and Westcott, Lewi and Harris, Honor and Windhorst, Rogier A and Holwerda, Benne W and Broadhurst, Thomas J and Coe, Dan and Cohen, Seth H and Du, Xiaojing and Driver, Simon P and Frye, Brenda and Grogin, Norman A and Hathi, Nimish P and Jansen, Rolf A and Koekemoer, Anton M and Marshall, Madeline A and Nonino, Mario and Ortiz III, Rafael and Pirzkal, Nor and Robotham, Aaron and Ryan, Russell E and Summers, Jake and D’Silva, Jordan C J and Willmer, Christopher N A and Yan, Haojing},
	month = may,
	year = {2025},
	pages = {774--805},
}

@article{patton_interacting_2024,
	title = {Interacting galaxies in the {IllustrisTNG} simulations – {VI}: {Reconstructed} orbits, close encounters, and mergers},
	volume = {529},
	copyright = {https://creativecommons.org/licenses/by/4.0/},
	issn = {0035-8711, 1365-2966},
	shorttitle = {Interacting galaxies in the {IllustrisTNG} simulations – {VI}},
	url = {https://academic.oup.com/mnras/article/529/2/1493/7616091},
	doi = {10.1093/mnras/stae608},
	language = {en},
	number = {2},
	urldate = {2026-03-03},
	journal = {Monthly Notices of the Royal Astronomical Society},
	author = {Patton, David R and Faria, Lawrence and Hani, Maan H and Torrey, Paul and Ellison, Sara L and Thakur, Shivani D and Westlake, Raven I},
	month = mar,
	year = {2024},
	pages = {1493--1506},
}

@article{conselice_direct_2022,
	title = {A {Direct} {Measurement} of {Galaxy} {Major} and {Minor} {Merger} {Rates} and {Stellar} {Mass} {Accretion} {Histories} at {Z} {\textless} 3 {Using} {Galaxy} {Pairs} in the {REFINE} {Survey}},
	volume = {940},
	issn = {0004-637X, 1538-4357},
	url = {https://iopscience.iop.org/article/10.3847/1538-4357/ac9b1a},
	doi = {10.3847/1538-4357/ac9b1a},
	number = {2},
	urldate = {2026-03-03},
	journal = {The Astrophysical Journal},
	author = {Conselice, Christopher J. and Mundy, Carl J. and Ferreira, Leonardo and Duncan, Kenneth},
	month = dec,
	year = {2022},
	pages = {168},
}

@article{mantha_major_2018,
	title = {Major merging history in {CANDELS}. {I}. {Evolution} of the incidence of massive galaxy–galaxy pairs from z = 3 to z ∼ 0},
	volume = {475},
	issn = {0035-8711, 1365-2966},
	url = {http://academic.oup.com/mnras/article/475/2/1549/4768277},
	doi = {10.1093/mnras/stx3260},
	language = {en},
	number = {2},
	urldate = {2026-03-03},
	journal = {Monthly Notices of the Royal Astronomical Society},
	author = {Mantha, Kameswara Bharadwaj and McIntosh, Daniel H and Brennan, Ryan and Ferguson, Henry C and Kodra, Dritan and Newman, Jeffrey A and Rafelski, Marc and Somerville, Rachel S and Conselice, Christopher J and Cook, Joshua S and Hathi, Nimish P and Koo, David C and Lotz, Jennifer M and Simmons, Brooke D and Straughn, Amber N and Snyder, Gregory F and Wuyts, Stijn and Bell, Eric F and Dekel, Avishai and Kartaltepe, Jeyhan and Kocevski, Dale D and Koekemoer, Anton M and Lee, Seong-Kook and Lucas, Ray A and Pacifici, Camilla and Peth, Michael A and Barro, Guillermo and Dahlen, Tomas and Finkelstein, Steven L and Fontana, Adriano and Galametz, Audrey and Grogin, Norman A and Guo, Yicheng and Mobasher, Bahram and Nayyeri, Hooshang and Pérez-González, Pablo G and Pforr, Janine and Santini, Paola and Stefanon, Mauro and Wiklind, Tommy},
	month = apr,
	year = {2018},
	pages = {1549--1573},
}

@article{mundy_consistent_2017,
	title = {A consistent measure of the merger histories of massive galaxies using close-pair statistics – {I}. {Major} mergers at z {\textless} 3.5},
	volume = {470},
	issn = {0035-8711, 1365-2966},
	url = {https://academic.oup.com/mnras/article-lookup/doi/10.1093/mnras/stx1238},
	doi = {10.1093/mnras/stx1238},
	language = {en},
	number = {3},
	urldate = {2026-03-03},
	journal = {Monthly Notices of the Royal Astronomical Society},
	author = {Mundy, Carl J. and Conselice, Christopher J. and Duncan, Kenneth J. and Almaini, Omar and Häußler, Boris and Hartley, William G.},
	month = sep,
	year = {2017},
	pages = {3507--3531},
}

@article{man_resolving_2016,
	title = {{RESOLVING} {THE} {DISCREPANCY} {OF} {GALAXY} {MERGER} {FRACTION} {MEASUREMENTS} {AT} z ∼ 0–3},
	volume = {830},
	issn = {0004-637X, 1538-4357},
	url = {https://iopscience.iop.org/article/10.3847/0004-637X/830/2/89},
	doi = {10.3847/0004-637X/830/2/89},
	number = {2},
	urldate = {2026-03-03},
	journal = {The Astrophysical Journal},
	author = {Man, Allison W. S. and Zirm, Andrew W. and Toft, Sune},
	month = oct,
	year = {2016},
	pages = {89},
}

@article{lopez-sanjuan_alhambra_2015,
	title = {The {ALHAMBRA} survey: accurate merger fractions derived by {PDF} analysis of photometrically close pairs},
	volume = {576},
	issn = {0004-6361, 1432-0746},
	shorttitle = {The {ALHAMBRA} survey},
	url = {http://www.aanda.org/10.1051/0004-6361/201424913},
	doi = {10.1051/0004-6361/201424913},
	urldate = {2026-03-03},
	journal = {Astronomy \& Astrophysics},
	author = {López-Sanjuan, C. and Cenarro, A. J. and Varela, J. and Viironen, K. and Molino, A. and Benítez, N. and Arnalte-Mur, P. and Ascaso, B. and Díaz-García, L. A. and Fernández-Soto, A. and Jiménez-Teja, Y. and Márquez, I. and Masegosa, J. and Moles, M. and Pović, M. and Aguerri, J. A. L. and Alfaro, E. and Aparicio-Villegas, T. and Broadhurst, T. and Cabrera-Caño, J. and Castander, F. J. and Cepa, J. and Cerviño, M. and Cristóbal-Hornillos, D. and Del Olmo, A. and González Delgado, R. M. and Husillos, C. and Infante, L. and Martínez, V. J. and Perea, J. and Prada, F. and Quintana, J. M.},
	month = apr,
	year = {2015},
	pages = {A53},
}

@article{ownsworth_minor_2014,
	title = {Minor versus major mergers: the stellar mass growth of massive galaxies from z = 3 using number density selection techniques},
	volume = {445},
	issn = {1365-2966, 0035-8711},
	shorttitle = {Minor versus major mergers},
	url = {http://academic.oup.com/mnras/article/445/3/2198/1037871/Minor-versus-major-mergers-the-stellar-mass-growth},
	doi = {10.1093/mnras/stu1802},
	language = {en},
	number = {3},
	urldate = {2026-03-03},
	journal = {Monthly Notices of the Royal Astronomical Society},
	author = {Ownsworth, Jamie R. and Conselice, Christopher J. and Mortlock, Alice and Hartley, William G. and Almaini, Omar and Duncan, Ken and Mundy, Carl J.},
	month = dec,
	year = {2014},
	pages = {2198--2213},
}

@article{bluck_structures_2012,
	title = {{THE} {STRUCTURES} {AND} {TOTAL} ({MINOR} + {MAJOR}) {MERGER} {HISTORIES} {OF} {MASSIVE} {GALAXIES} {UP} {TO} \textit{z} ∼ 3 {IN} {THE} \textit{{HST}} {GOODS} {NICMOS} {SURVEY}: {A} {POSSIBLE} {SOLUTION} {TO} {THE} {SIZE} {EVOLUTION} {PROBLEM}},
	volume = {747},
	issn = {0004-637X, 1538-4357},
	shorttitle = {{THE} {STRUCTURES} {AND} {TOTAL} ({MINOR} + {MAJOR}) {MERGER} {HISTORIES} {OF} {MASSIVE} {GALAXIES} {UP} {TO} \textit{z} ∼ 3 {IN} {THE} \textit{{HST}} {GOODS} {NICMOS} {SURVEY}},
	url = {https://iopscience.iop.org/article/10.1088/0004-637X/747/1/34},
	doi = {10.1088/0004-637X/747/1/34},
	number = {1},
	urldate = {2026-03-03},
	journal = {The Astrophysical Journal},
	author = {Bluck, Asa F. L. and Conselice, Christopher J. and Buitrago, Fernando and Grützbauch, Ruth and Hoyos, Carlos and Mortlock, Alice and Bauer, Amanda E.},
	month = mar,
	year = {2012},
	pages = {34},
}

@article{lotz_major_2011,
	title = {{THE} {MAJOR} {AND} {MINOR} {GALAXY} {MERGER} {RATES} {AT} \textit{z} {\textless} 1.5},
	volume = {742},
	issn = {0004-637X, 1538-4357},
	url = {https://iopscience.iop.org/article/10.1088/0004-637X/742/2/103},
	doi = {10.1088/0004-637X/742/2/103},
	number = {2},
	urldate = {2026-03-03},
	journal = {The Astrophysical Journal},
	author = {Lotz, Jennifer M. and Jonsson, Patrik and Cox, T. J. and Croton, Darren and Primack, Joel R. and Somerville, Rachel S. and Stewart, Kyle},
	month = dec,
	year = {2011},
	pages = {103},
}

@article{conselice_structures_2008,
	title = {The structures of distant galaxies – {I}. {Galaxy} structures and the merger rate to z ∼ 3 in the {Hubble} {Ultra}-{Deep} {Field}},
	volume = {386},
	issn = {0035-8711, 1365-2966},
	url = {https://academic.oup.com/mnras/article-lookup/doi/10.1111/j.1365-2966.2008.13069.x},
	doi = {10.1111/j.1365-2966.2008.13069.x},
	language = {en},
	number = {2},
	urldate = {2026-03-03},
	journal = {Monthly Notices of the Royal Astronomical Society},
	author = {Conselice, Christopher J. and Rajgor, Sheena and Myers, Robert},
	month = may,
	year = {2008},
	pages = {909--927},
}

@article{bernet_dark_2025,
	title = {Dark matter spiral arms in {Milky} {Way}-like halos},
	volume = {697},
	copyright = {https://creativecommons.org/licenses/by/4.0},
	issn = {0004-6361, 1432-0746},
	url = {https://www.aanda.org/10.1051/0004-6361/202554458},
	doi = {10.1051/0004-6361/202554458},
	urldate = {2026-03-03},
	journal = {Astronomy \& Astrophysics},
	author = {Bernet, Marcel and Ramos, Pau and Antoja, Teresa and Debattista, Victor P. and Weinberg, Martin D. and Amarante, João A. S. and Grand, Robert J. J. and Jiménez-Arranz, Óscar and Laporte, Chervin F. P. and Petersen, Michael S. and Roca-Fàbrega, Santi and Romero-Gómez, Mercè},
	month = may,
	year = {2025},
	pages = {A214},
}

@article{gnedin_response_2004,
	title = {Response of {Dark} {Matter} {Halos} to {Condensation} of {Baryons}: {Cosmological} {Simulations} and {Improved} {Adiabatic} {Contraction} {Model}},
	volume = {616},
	issn = {0004-637X, 1538-4357},
	shorttitle = {Response of {Dark} {Matter} {Halos} to {Condensation} of {Baryons}},
	url = {https://iopscience.iop.org/article/10.1086/424914},
	doi = {10.1086/424914},
	language = {en},
	number = {1},
	urldate = {2026-03-03},
	journal = {The Astrophysical Journal},
	author = {Gnedin, Oleg Y. and Kravtsov, Andrey V. and Klypin, Anatoly A. and Nagai, Daisuke},
	month = nov,
	year = {2004},
	pages = {16--26},
}

@article{dillamore_merger-induced_2022,
	title = {Merger-induced galaxy transformations in the {\textless}span style="font-variant:small-caps;"{\textgreater}artemis{\textless}/span{\textgreater} simulations},
	volume = {513},
	copyright = {https://academic.oup.com/journals/pages/open\_access/funder\_policies/chorus/standard\_publication\_model},
	issn = {0035-8711, 1365-2966},
	shorttitle = {Merger-induced galaxy transformations in the {\textless}span style="font-variant},
	url = {https://academic.oup.com/mnras/article/513/2/1867/6568563},
	doi = {10.1093/mnras/stac1038},
	language = {en},
	number = {2},
	urldate = {2026-03-03},
	journal = {Monthly Notices of the Royal Astronomical Society},
	author = {Dillamore, Adam M and Belokurov, Vasily and Font, Andreea S and McCarthy, Ian G},
	month = may,
	year = {2022},
	pages = {1867--1886},
}

@article{ash_stellar_2024,
	title = {Stellar {Bars} {Form} {Dark} {Matter} {Counterparts} in {TNG50}},
	volume = {976},
	issn = {0004-637X, 1538-4357},
	url = {https://iopscience.iop.org/article/10.3847/1538-4357/ad863a},
	doi = {10.3847/1538-4357/ad863a},
	number = {2},
	urldate = {2026-03-02},
	journal = {The Astrophysical Journal},
	author = {Ash, Neil and Valluri, Monica and Chen, Yingtian and Bell, Eric F.},
	month = dec,
	year = {2024},
	pages = {189},
}

@misc{darragh-ford_shaping_2025,
	title = {Shaping the {Milky} {Way}. {II}. {The} dark matter halo's response to the {LMC}'s passage in a cosmological context},
	copyright = {Creative Commons Attribution 4.0 International},
	url = {https://arxiv.org/abs/2511.02031},
	doi = {10.48550/ARXIV.2511.02031},
	urldate = {2026-02-17},
	publisher = {arXiv},
	author = {Darragh-Ford, Elise and Garavito-Camargo, Nicolas and Arora, Arpit and Wechsler, Risa H. and Mansfield, Phil and Besla, Gurtina and Petersen, Michael S. and Weinberg, Martin D. and Varela-Lavin, Silvio and Buch, Deveshi and Cunningham, Emily C. and Daniel, Kathryne J. and Gomez, Facundo A. and Johnston, Kathryn V. and Laporte, Chervin F. P. and Mao, Yao-Yuan and Nadler, Ethan O. and Sanderson, Robyn},
	year = {2025},
	note = {Version Number: 1},
}

@misc{boyle_spherical_2023,
	title = {The spherical package},
	copyright = {MIT License},
	url = {https://zenodo.org/doi/10.5281/zenodo.10214833},
	doi = {10.5281/ZENODO.10214833},
	urldate = {2026-01-21},
	publisher = {Zenodo},
	author = {Boyle, Michael},
	month = nov,
	year = {2023},
}

@misc{boyle_quaternionic_2024,
	title = {The quaternionic package},
	copyright = {MIT License},
	url = {https://zenodo.org/doi/10.5281/zenodo.13350880},
	doi = {10.5281/ZENODO.13350880},
	urldate = {2026-01-21},
	publisher = {Zenodo},
	author = {Boyle, Michael},
	month = aug,
	year = {2024},
}

@article{schwarzschild_numerical_1979,
	title = {A numerical model for a triaxial stellar system in dynamical equilibrium},
	volume = {232},
	issn = {0004-637X, 1538-4357},
	url = {http://adsabs.harvard.edu/doi/10.1086/157282},
	doi = {10.1086/157282},
	language = {en},
	urldate = {2025-09-04},
	journal = {The Astrophysical Journal},
	author = {Schwarzschild, M.},
	month = aug,
	year = {1979},
	pages = {236},
}

@article{hernquist_analytical_1990,
	title = {An analytical model for spherical galaxies and bulges},
	volume = {356},
	issn = {0004-637X, 1538-4357},
	url = {http://adsabs.harvard.edu/doi/10.1086/168845},
	doi = {10.1086/168845},
	language = {en},
	urldate = {2025-09-04},
	journal = {The Astrophysical Journal},
	author = {Hernquist, Lars},
	month = jun,
	year = {1990},
	pages = {359},
}

@article{pillepich_milky_2024,
	title = {Milky {Way} and {Andromeda} analogues from the {TNG50} simulation},
	volume = {535},
	copyright = {https://creativecommons.org/licenses/by/4.0/},
	issn = {0035-8711, 1365-2966},
	url = {https://academic.oup.com/mnras/article/535/2/1721/7760400},
	doi = {10.1093/mnras/stae2165},
	language = {en},
	number = {2},
	urldate = {2025-09-04},
	journal = {Monthly Notices of the Royal Astronomical Society},
	author = {Pillepich, Annalisa and Sotillo-Ramos, Diego and Ramesh, Rahul and Nelson, Dylan and Engler, Christoph and Rodriguez-Gomez, Vicente and Fournier, Martin and Donnari, Martina and Springel, Volker and Hernquist, Lars},
	month = nov,
	year = {2024},
	pages = {1721--1762},
}

@article{nelson_illustristng_2019,
	title = {The {IllustrisTNG} simulations: public data release},
	volume = {6},
	issn = {2197-7909},
	shorttitle = {The {IllustrisTNG} simulations},
	url = {https://link.springer.com/10.1186/s40668-019-0028-x},
	doi = {10.1186/s40668-019-0028-x},
	language = {en},
	number = {1},
	urldate = {2025-09-04},
	journal = {Computational Astrophysics and Cosmology},
	author = {Nelson, Dylan and Springel, Volker and Pillepich, Annalisa and Rodriguez-Gomez, Vicente and Torrey, Paul and Genel, Shy and Vogelsberger, Mark and Pakmor, Ruediger and Marinacci, Federico and Weinberger, Rainer and Kelley, Luke and Lovell, Mark and Diemer, Benedikt and Hernquist, Lars},
	month = dec,
	year = {2019},
	pages = {2},
}

@article{arora_shaping_2025,
	title = {Shaping the {Milky} {Way}: {The} {Interplay} of {Mergers} and {Cosmic} {Filaments}},
	volume = {988},
	issn = {0004-637X, 1538-4357},
	shorttitle = {Shaping the {Milky} {Way}},
	url = {https://iopscience.iop.org/article/10.3847/1538-4357/ade30d},
	doi = {10.3847/1538-4357/ade30d},
	number = {2},
	urldate = {2025-08-28},
	journal = {The Astrophysical Journal},
	author = {Arora, Arpit and Garavito-Camargo, Nicolás and Sanderson, Robyn E. and Weinberg, Martin D. and Petersen, Michael S. and Varela-Lavin, Silvio and Gómez, Facundo A. and Johnston, Kathryn V. and Laporte, Chervin F. P. and Shipp, Nora and Hunt, Jason A. S. and Besla, Gurtina and Darragh-Ford, Elise and Panithanpaisal, Nondh and Daniel, Kathryne J. and {The EXP collaboration}},
	month = aug,
	year = {2025},
	pages = {190},
}

@article{collier_coupling_2021,
	title = {The {Coupling} of {Galactic} {Dark} {Matter} {Halos} with {Stellar} {Bars}},
	volume = {915},
	issn = {0004-637X, 1538-4357},
	url = {https://iopscience.iop.org/article/10.3847/1538-4357/ac004d},
	doi = {10.3847/1538-4357/ac004d},
	number = {1},
	urldate = {2024-06-20},
	journal = {The Astrophysical Journal},
	author = {Collier, Angela and Madigan, Ann-Marie},
	month = jul,
	year = {2021},
	pages = {23},
}

@article{marostica_response_2024,
	title = {The {Response} of the {Inner} {Dark} {Matter} {Halo} to {Stellar} {Bars}},
	volume = {12},
	copyright = {https://creativecommons.org/licenses/by/4.0/},
	issn = {2075-4434},
	url = {https://www.mdpi.com/2075-4434/12/3/27},
	doi = {10.3390/galaxies12030027},
	language = {en},
	number = {3},
	urldate = {2024-05-30},
	journal = {Galaxies},
	author = {Marostica, Daniel A. and Machado, Rubens E. G. and Athanassoula, E. and Manos, T.},
	month = may,
	year = {2024},
	pages = {27},
}

@book{binney_galactic_2008,
	address = {Princeton},
	edition = {2nd ed},
	series = {Princeton series in astrophysics},
	title = {Galactic dynamics},
	isbn = {978-0-691-13026-2 978-0-691-13027-9},
	publisher = {Princeton University Press},
	author = {Binney, James and Tremaine, Scott},
	year = {2008},
}

@article{ash_figure_2023,
	title = {Figure {Rotation} of {IllustrisTNG} {Halos}},
	volume = {955},
	issn = {0004-637X, 1538-4357},
	url = {https://iopscience.iop.org/article/10.3847/1538-4357/acf30c},
	doi = {10.3847/1538-4357/acf30c},
	number = {2},
	urldate = {2023-11-20},
	journal = {The Astrophysical Journal},
	author = {Ash, Neil and Valluri, Monica},
	month = oct,
	year = {2023},
	pages = {111},
}

@article{prada_dark_2019,
	title = {Dark matter halo shapes in the {Auriga} simulations},
	volume = {490},
	issn = {0035-8711, 1365-2966},
	url = {https://academic.oup.com/mnras/article/490/4/4877/5586599},
	doi = {10.1093/mnras/stz2873},
	language = {en},
	number = {4},
	urldate = {2023-04-21},
	journal = {Monthly Notices of the Royal Astronomical Society},
	author = {Prada, Jesus and Forero-Romero, Jaime E and Grand, Robert J J and Pakmor, Rüdiger and Springel, Volker},
	month = dec,
	year = {2019},
	pages = {4877--4888},
}

@article{peebles_origin_1969,
	title = {Origin of the {Angular} {Momentum} of {Galaxies}},
	volume = {155},
	issn = {0004-637X, 1538-4357},
	url = {http://adsabs.harvard.edu/doi/10.1086/149876},
	doi = {10.1086/149876},
	language = {en},
	urldate = {2023-04-06},
	journal = {The Astrophysical Journal},
	author = {Peebles, P. J. E.},
	month = feb,
	year = {1969},
	pages = {393},
}

@article{kazantzidis_effect_2004,
	title = {The {Effect} of {Gas} {Cooling} on the {Shapes} of {Dark} {Matter} {Halos}},
	volume = {611},
	issn = {0004-637X, 1538-4357},
	url = {https://iopscience.iop.org/article/10.1086/423992},
	doi = {10.1086/423992},
	language = {en},
	number = {2},
	urldate = {2023-04-06},
	journal = {The Astrophysical Journal},
	author = {Kazantzidis, Stelios and Kravtsov, Andrey V. and Zentner, Andrew R. and Allgood, Brandon and Nagai, Daisuke and Moore, Ben},
	month = aug,
	year = {2004},
	pages = {L73--L76},
}

@article{bullock_profiles_2001,
	title = {Profiles of dark haloes: evolution, scatter and environment},
	volume = {321},
	issn = {0035-8711},
	shorttitle = {Profiles of dark haloes},
	url = {https://ui.adsabs.harvard.edu/abs/2001MNRAS.321..559B},
	doi = {10.1046/j.1365-8711.2001.04068.x},
	urldate = {2023-04-05},
	journal = {Monthly Notices of the Royal Astronomical Society},
	author = {Bullock, J. S. and Kolatt, T. S. and Sigad, Y. and Somerville, R. S. and Kravtsov, A. V. and Klypin, A. A. and Primack, J. R. and Dekel, A.},
	month = mar,
	year = {2001},
	note = {ADS Bibcode: 2001MNRAS.321..559B},
	pages = {559--575},
}

@article{planck_collaboration_planck_2016,
	title = {\textit{{Planck}} 2015 results: {XIII}. {Cosmological} parameters},
	volume = {594},
	issn = {0004-6361, 1432-0746},
	shorttitle = {\textit{{Planck}} 2015 results},
	url = {http://www.aanda.org/10.1051/0004-6361/201525830},
	doi = {10.1051/0004-6361/201525830},
	urldate = {2023-04-05},
	journal = {Astronomy \& Astrophysics},
	author = {{Planck Collaboration} and Ade, P. A. R. and Aghanim, N. and Arnaud, M. and Ashdown, M. and Aumont, J. and Baccigalupi, C. and Banday, A. J. and Barreiro, R. B. and Bartlett, J. G. and Bartolo, N. and Battaner, E. and Battye, R. and Benabed, K. and Benoît, A. and Benoit-Lévy, A. and Bernard, J.-P. and Bersanelli, M. and Bielewicz, P. and Bock, J. J. and Bonaldi, A. and Bonavera, L. and Bond, J. R. and Borrill, J. and Bouchet, F. R. and Boulanger, F. and Bucher, M. and Burigana, C. and Butler, R. C. and Calabrese, E. and Cardoso, J.-F. and Catalano, A. and Challinor, A. and Chamballu, A. and Chary, R.-R. and Chiang, H. C. and Chluba, J. and Christensen, P. R. and Church, S. and Clements, D. L. and Colombi, S. and Colombo, L. P. L. and Combet, C. and Coulais, A. and Crill, B. P. and Curto, A. and Cuttaia, F. and Danese, L. and Davies, R. D. and Davis, R. J. and de Bernardis, P. and de Rosa, A. and de Zotti, G. and Delabrouille, J. and Désert, F.-X. and Di Valentino, E. and Dickinson, C. and Diego, J. M. and Dolag, K. and Dole, H. and Donzelli, S. and Doré, O. and Douspis, M. and Ducout, A. and Dunkley, J. and Dupac, X. and Efstathiou, G. and Elsner, F. and Enßlin, T. A. and Eriksen, H. K. and Farhang, M. and Fergusson, J. and Finelli, F. and Forni, O. and Frailis, M. and Fraisse, A. A. and Franceschi, E. and Frejsel, A. and Galeotta, S. and Galli, S. and Ganga, K. and Gauthier, C. and Gerbino, M. and Ghosh, T. and Giard, M. and Giraud-Héraud, Y. and Giusarma, E. and Gjerløw, E. and González-Nuevo, J. and Górski, K. M. and Gratton, S. and Gregorio, A. and Gruppuso, A. and Gudmundsson, J. E. and Hamann, J. and Hansen, F. K. and Hanson, D. and Harrison, D. L. and Helou, G. and Henrot-Versillé, S. and Hernández-Monteagudo, C. and Herranz, D. and Hildebrandt, S. R. and Hivon, E. and Hobson, M. and Holmes, W. A. and Hornstrup, A. and Hovest, W. and Huang, Z. and Huffenberger, K. M. and Hurier, G. and Jaffe, A. H. and Jaffe, T. R. and Jones, W. C. and Juvela, M. and Keihänen, E. and Keskitalo, R. and Kisner, T. S. and Kneissl, R. and Knoche, J. and Knox, L. and Kunz, M. and Kurki-Suonio, H. and Lagache, G. and Lähteenmäki, A. and Lamarre, J.-M. and Lasenby, A. and Lattanzi, M. and Lawrence, C. R. and Leahy, J. P. and Leonardi, R. and Lesgourgues, J. and Levrier, F. and Lewis, A. and Liguori, M. and Lilje, P. B. and Linden-Vørnle, M. and López-Caniego, M. and Lubin, P. M. and Macías-Pérez, J. F. and Maggio, G. and Maino, D. and Mandolesi, N. and Mangilli, A. and Marchini, A. and Maris, M. and Martin, P. G. and Martinelli, M. and Martínez-González, E. and Masi, S. and Matarrese, S. and McGehee, P. and Meinhold, P. R. and Melchiorri, A. and Melin, J.-B. and Mendes, L. and Mennella, A. and Migliaccio, M. and Millea, M. and Mitra, S. and Miville-Deschênes, M.-A. and Moneti, A. and Montier, L. and Morgante, G. and Mortlock, D. and Moss, A. and Munshi, D. and Murphy, J. A. and Naselsky, P. and Nati, F. and Natoli, P. and Netterfield, C. B. and Nørgaard-Nielsen, H. U. and Noviello, F. and Novikov, D. and Novikov, I. and Oxborrow, C. A. and Paci, F. and Pagano, L. and Pajot, F. and Paladini, R. and Paoletti, D. and Partridge, B. and Pasian, F. and Patanchon, G. and Pearson, T. J. and Perdereau, O. and Perotto, L. and Perrotta, F. and Pettorino, V. and Piacentini, F. and Piat, M. and Pierpaoli, E. and Pietrobon, D. and Plaszczynski, S. and Pointecouteau, E. and Polenta, G. and Popa, L. and Pratt, G. W. and Prézeau, G. and Prunet, S. and Puget, J.-L. and Rachen, J. P. and Reach, W. T. and Rebolo, R. and Reinecke, M. and Remazeilles, M. and Renault, C. and Renzi, A. and Ristorcelli, I. and Rocha, G. and Rosset, C. and Rossetti, M. and Roudier, G. and Rouillé d’Orfeuil, B. and Rowan-Robinson, M. and Rubiño-Martín, J. A. and Rusholme, B. and Said, N. and Salvatelli, V. and Salvati, L. and Sandri, M. and Santos, D. and Savelainen, M. and Savini, G. and Scott, D. and Seiffert, M. D. and Serra, P. and Shellard, E. P. S. and Spencer, L. D. and Spinelli, M. and Stolyarov, V. and Stompor, R. and Sudiwala, R. and Sunyaev, R. and Sutton, D. and Suur-Uski, A.-S. and Sygnet, J.-F. and Tauber, J. A. and Terenzi, L. and Toffolatti, L. and Tomasi, M. and Tristram, M. and Trombetti, T. and Tucci, M. and Tuovinen, J. and Türler, M. and Umana, G. and Valenziano, L. and Valiviita, J. and Van Tent, F. and Vielva, P. and Villa, F. and Wade, L. A. and Wandelt, B. D. and Wehus, I. K. and White, M. and White, S. D. M. and Wilkinson, A. and Yvon, D. and Zacchei, A. and Zonca, A.},
	month = oct,
	year = {2016},
	pages = {A13},
}

@article{the_astropy_collaboration_astropy_2018,
	title = {The {Astropy} {Project}: {Building} an {Open}-science {Project} and {Status} of the v2.0 {Core} {Package}},
	volume = {156},
	issn = {1538-3881},
	shorttitle = {The {Astropy} {Project}},
	url = {https://iopscience.iop.org/article/10.3847/1538-3881/aabc4f},
	doi = {10.3847/1538-3881/aabc4f},
	number = {3},
	urldate = {2023-03-01},
	journal = {The Astronomical Journal},
	author = {{The Astropy Collaboration} and Price-Whelan, A. M. and Sipőcz, B. M. and Günther, H. M. and Lim, P. L. and Crawford, S. M. and Conseil, S. and Shupe, D. L. and Craig, M. W. and Dencheva, N. and Ginsburg, A. and VanderPlas, J. T. and Bradley, L. D. and Pérez-Suárez, D. and de Val-Borro, M. and {(Primary Paper Contributors)} and Aldcroft, T. L. and Cruz, K. L. and Robitaille, T. P. and Tollerud, E. J. and {(Astropy Coordination Committee)} and Ardelean, C. and Babej, T. and Bach, Y. P. and Bachetti, M. and Bakanov, A. V. and Bamford, S. P. and Barentsen, G. and Barmby, P. and Baumbach, A. and Berry, K. L. and Biscani, F. and Boquien, M. and Bostroem, K. A. and Bouma, L. G. and Brammer, G. B. and Bray, E. M. and Breytenbach, H. and Buddelmeijer, H. and Burke, D. J. and Calderone, G. and Rodríguez, J. L. Cano and Cara, M. and Cardoso, J. V. M. and Cheedella, S. and Copin, Y. and Corrales, L. and Crichton, D. and D’Avella, D. and Deil, C. and Depagne, É. and Dietrich, J. P. and Donath, A. and Droettboom, M. and Earl, N. and Erben, T. and Fabbro, S. and Ferreira, L. A. and Finethy, T. and Fox, R. T. and Garrison, L. H. and Gibbons, S. L. J. and Goldstein, D. A. and Gommers, R. and Greco, J. P. and Greenfield, P. and Groener, A. M. and Grollier, F. and Hagen, A. and Hirst, P. and Homeier, D. and Horton, A. J. and Hosseinzadeh, G. and Hu, L. and Hunkeler, J. S. and Ivezić, Ž. and Jain, A. and Jenness, T. and Kanarek, G. and Kendrew, S. and Kern, N. S. and Kerzendorf, W. E. and Khvalko, A. and King, J. and Kirkby, D. and Kulkarni, A. M. and Kumar, A. and Lee, A. and Lenz, D. and Littlefair, S. P. and Ma, Z. and Macleod, D. M. and Mastropietro, M. and McCully, C. and Montagnac, S. and Morris, B. M. and Mueller, M. and Mumford, S. J. and Muna, D. and Murphy, N. A. and Nelson, S. and Nguyen, G. H. and Ninan, J. P. and Nöthe, M. and Ogaz, S. and Oh, S. and Parejko, J. K. and Parley, N. and Pascual, S. and Patil, R. and Patil, A. A. and Plunkett, A. L. and Prochaska, J. X. and Rastogi, T. and Janga, V. Reddy and Sabater, J. and Sakurikar, P. and Seifert, M. and Sherbert, L. E. and Sherwood-Taylor, H. and Shih, A. Y. and Sick, J. and Silbiger, M. T. and Singanamalla, S. and Singer, L. P. and Sladen, P. H. and Sooley, K. A. and Sornarajah, S. and Streicher, O. and Teuben, P. and Thomas, S. W. and Tremblay, G. R. and Turner, J. E. H. and Terrón, V. and Kerkwijk, M. H. van and de la Vega, A. and Watkins, L. L. and Weaver, B. A. and Whitmore, J. B. and Woillez, J. and Zabalza, V. and {(Astropy Contributors)}},
	month = aug,
	year = {2018},
	pages = {123},
}

@article{virtanen_scipy_2020,
	title = {{SciPy} 1.0: fundamental algorithms for scientific computing in {Python}},
	volume = {17},
	issn = {1548-7091, 1548-7105},
	shorttitle = {{SciPy} 1.0},
	url = {http://www.nature.com/articles/s41592-019-0686-2},
	doi = {10.1038/s41592-019-0686-2},
	language = {en},
	number = {3},
	urldate = {2023-03-01},
	journal = {Nature Methods},
	author = {Virtanen, Pauli and Gommers, Ralf and Oliphant, Travis E. and Haberland, Matt and Reddy, Tyler and Cournapeau, David and Burovski, Evgeni and Peterson, Pearu and Weckesser, Warren and Bright, Jonathan and van der Walt, Stéfan J. and Brett, Matthew and Wilson, Joshua and Millman, K. Jarrod and Mayorov, Nikolay and Nelson, Andrew R. J. and Jones, Eric and Kern, Robert and Larson, Eric and Carey, C J and Polat, İlhan and Feng, Yu and Moore, Eric W. and VanderPlas, Jake and Laxalde, Denis and Perktold, Josef and Cimrman, Robert and Henriksen, Ian and Quintero, E. A. and Harris, Charles R. and Archibald, Anne M. and Ribeiro, Antônio H. and Pedregosa, Fabian and van Mulbregt, Paul and {SciPy 1.0 Contributors} and Vijaykumar, Aditya and Bardelli, Alessandro Pietro and Rothberg, Alex and Hilboll, Andreas and Kloeckner, Andreas and Scopatz, Anthony and Lee, Antony and Rokem, Ariel and Woods, C. Nathan and Fulton, Chad and Masson, Charles and Häggström, Christian and Fitzgerald, Clark and Nicholson, David A. and Hagen, David R. and Pasechnik, Dmitrii V. and Olivetti, Emanuele and Martin, Eric and Wieser, Eric and Silva, Fabrice and Lenders, Felix and Wilhelm, Florian and Young, G. and Price, Gavin A. and Ingold, Gert-Ludwig and Allen, Gregory E. and Lee, Gregory R. and Audren, Hervé and Probst, Irvin and Dietrich, Jörg P. and Silterra, Jacob and Webber, James T and Slavič, Janko and Nothman, Joel and Buchner, Johannes and Kulick, Johannes and Schönberger, Johannes L. and de Miranda Cardoso, José Vinícius and Reimer, Joscha and Harrington, Joseph and Rodríguez, Juan Luis Cano and Nunez-Iglesias, Juan and Kuczynski, Justin and Tritz, Kevin and Thoma, Martin and Newville, Matthew and Kümmerer, Matthias and Bolingbroke, Maximilian and Tartre, Michael and Pak, Mikhail and Smith, Nathaniel J. and Nowaczyk, Nikolai and Shebanov, Nikolay and Pavlyk, Oleksandr and Brodtkorb, Per A. and Lee, Perry and McGibbon, Robert T. and Feldbauer, Roman and Lewis, Sam and Tygier, Sam and Sievert, Scott and Vigna, Sebastiano and Peterson, Stefan and More, Surhud and Pudlik, Tadeusz and Oshima, Takuya and Pingel, Thomas J. and Robitaille, Thomas P. and Spura, Thomas and Jones, Thouis R. and Cera, Tim and Leslie, Tim and Zito, Tiziano and Krauss, Tom and Upadhyay, Utkarsh and Halchenko, Yaroslav O. and Vázquez-Baeza, Yoshiki},
	month = mar,
	year = {2020},
	pages = {261--272},
}

@article{harris_array_2020,
	title = {Array programming with {NumPy}},
	volume = {585},
	issn = {0028-0836, 1476-4687},
	url = {https://www.nature.com/articles/s41586-020-2649-2},
	doi = {10.1038/s41586-020-2649-2},
	language = {en},
	number = {7825},
	urldate = {2023-03-01},
	journal = {Nature},
	author = {Harris, Charles R. and Millman, K. Jarrod and van der Walt, Stéfan J. and Gommers, Ralf and Virtanen, Pauli and Cournapeau, David and Wieser, Eric and Taylor, Julian and Berg, Sebastian and Smith, Nathaniel J. and Kern, Robert and Picus, Matti and Hoyer, Stephan and van Kerkwijk, Marten H. and Brett, Matthew and Haldane, Allan and del Río, Jaime Fernández and Wiebe, Mark and Peterson, Pearu and Gérard-Marchant, Pierre and Sheppard, Kevin and Reddy, Tyler and Weckesser, Warren and Abbasi, Hameer and Gohlke, Christoph and Oliphant, Travis E.},
	month = sep,
	year = {2020},
	pages = {357--362},
}

@article{vasiliev_agama_2019,
	title = {{AGAMA}: action-based galaxy modelling architecture},
	volume = {482},
	issn = {0035-8711, 1365-2966},
	shorttitle = {{AGAMA}},
	url = {https://academic.oup.com/mnras/article/482/2/1525/5114593},
	doi = {10.1093/mnras/sty2672},
	language = {en},
	number = {2},
	urldate = {2023-03-01},
	journal = {Monthly Notices of the Royal Astronomical Society},
	author = {Vasiliev, Eugene},
	month = jan,
	year = {2019},
	pages = {1525--1544},
}

@article{nelson_first_2019,
	title = {First results from the {TNG50} simulation: galactic outflows driven by supernovae and black hole feedback},
	volume = {490},
	issn = {0035-8711, 1365-2966},
	shorttitle = {First results from the {TNG50} simulation},
	url = {https://academic.oup.com/mnras/article/490/3/3234/5556547},
	doi = {10.1093/mnras/stz2306},
	language = {en},
	number = {3},
	urldate = {2023-02-27},
	journal = {Monthly Notices of the Royal Astronomical Society},
	author = {Nelson, Dylan and Pillepich, Annalisa and Springel, Volker and Pakmor, Rüdiger and Weinberger, Rainer and Genel, Shy and Torrey, Paul and Vogelsberger, Mark and Marinacci, Federico and Hernquist, Lars},
	month = dec,
	year = {2019},
	pages = {3234--3261},
}

@article{deibel_orbital_2011,
	title = {{THE} {ORBITAL} {STRUCTURE} {OF} {TRIAXIAL} {GALAXIES} {WITH} {FIGURE} {ROTATION}},
	volume = {728},
	issn = {0004-637X, 1538-4357},
	url = {https://iopscience.iop.org/article/10.1088/0004-637X/728/2/128},
	doi = {10.1088/0004-637X/728/2/128},
	number = {2},
	urldate = {2023-02-25},
	journal = {The Astrophysical Journal},
	author = {Deibel, Alex T. and Valluri, Monica and Merritt, David},
	month = feb,
	year = {2011},
	pages = {128},
}

@article{pillepich_first_2019,
	title = {First results from the {TNG50} simulation: the evolution of stellar and gaseous discs across cosmic time},
	volume = {490},
	issn = {0035-8711},
	shorttitle = {First results from the {TNG50} simulation},
	url = {https://ui.adsabs.harvard.edu/abs/2019MNRAS.490.3196P},
	doi = {10.1093/mnras/stz2338},
	urldate = {2023-02-14},
	journal = {Monthly Notices of the Royal Astronomical Society},
	author = {Pillepich, Annalisa and Nelson, Dylan and Springel, Volker and Pakmor, Rüdiger and Torrey, Paul and Weinberger, Rainer and Vogelsberger, Mark and Marinacci, Federico and Genel, Shy and van der Wel, Arjen and Hernquist, Lars},
	month = dec,
	year = {2019},
	note = {ADS Bibcode: 2019MNRAS.490.3196P},
	pages = {3196--3233},
}

@article{chua_shape_2019,
	title = {Shape of dark matter haloes in the {Illustris} simulation: effects of baryons},
	volume = {484},
	issn = {0035-8711},
	shorttitle = {Shape of dark matter haloes in the {Illustris} simulation},
	url = {https://doi.org/10.1093/mnras/sty3531},
	doi = {10.1093/mnras/sty3531},
	number = {1},
	urldate = {2023-01-31},
	journal = {Monthly Notices of the Royal Astronomical Society},
	author = {Chua, Kun Ting Eddie and Pillepich, Annalisa and Vogelsberger, Mark and Hernquist, Lars},
	month = mar,
	year = {2019},
	pages = {476--493},
}

@article{lagos_quantifying_2018,
	title = {Quantifying the impact of mergers on the angular momentum of simulated galaxies},
	volume = {473},
	issn = {0035-8711},
	url = {https://doi.org/10.1093/mnras/stx2667},
	doi = {10.1093/mnras/stx2667},
	number = {4},
	urldate = {2023-01-22},
	journal = {Monthly Notices of the Royal Astronomical Society},
	author = {Lagos, Claudia del P. and Stevens, Adam R. H. and Bower, Richard G. and Davis, Timothy A. and Contreras, Sergio and Padilla, Nelson D. and Obreschkow, Danail and Croton, Darren and Trayford, James W. and Welker, Charlotte and Theuns, Tom},
	month = feb,
	year = {2018},
	pages = {4956--4974},
}

@article{dehnen_very_2000,
	title = {A {Very} {Fast} and {Momentum}-conserving {Tree} {Code}},
	volume = {536},
	issn = {0004-637X},
	url = {https://ui.adsabs.harvard.edu/abs/2000ApJ...536L..39D},
	doi = {10.1086/312724},
	urldate = {2022-11-11},
	journal = {The Astrophysical Journal},
	author = {Dehnen, Walter},
	month = jun,
	year = {2000},
	note = {ADS Bibcode: 2000ApJ...536L..39D},
	pages = {L39--L42},
}

@article{dubinski_cosmological_1992,
	title = {Cosmological {Tidal} {Shear}},
	volume = {401},
	issn = {0004-637X},
	url = {https://ui.adsabs.harvard.edu/abs/1992ApJ...401..441D},
	doi = {10.1086/172076},
	urldate = {2022-10-18},
	journal = {The Astrophysical Journal},
	author = {Dubinski, John},
	month = dec,
	year = {1992},
	note = {ADS Bibcode: 1992ApJ...401..441D},
	pages = {441},
}

@article{garavito-camargo_quantifying_2021,
	title = {Quantifying the {Impact} of the {Large} {Magellanic} {Cloud} on the {Structure} of the {Milky} {Way}'s {Dark} {Matter} {Halo} {Using} {Basis} {Function} {Expansions}},
	volume = {919},
	issn = {0004-637X},
	url = {https://ui.adsabs.harvard.edu/abs/2021ApJ...919..109G},
	doi = {10.3847/1538-4357/ac0b44},
	urldate = {2022-10-18},
	journal = {The Astrophysical Journal},
	author = {Garavito-Camargo, Nicolás and Besla, Gurtina and Laporte, Chervin F. P. and Price-Whelan, Adrian M. and Cunningham, Emily C. and Johnston, Kathryn V. and Weinberg, Martin and Gómez, Facundo A.},
	month = oct,
	year = {2021},
	note = {ADS Bibcode: 2021ApJ...919..109G},
	pages = {109},
}

@article{valluri_detecting_2021,
	title = {Detecting the {Figure} {Rotation} of {Dark} {Matter} {Halos} with {Tidal} {Streams}},
	volume = {910},
	issn = {0004-637X},
	url = {https://ui.adsabs.harvard.edu/abs/2021ApJ...910..150V},
	doi = {10.3847/1538-4357/abe534},
	urldate = {2022-10-18},
	journal = {The Astrophysical Journal},
	author = {Valluri, Monica and Price-Whelan, Adrian M. and Snyder, Sarah J.},
	month = apr,
	year = {2021},
	note = {ADS Bibcode: 2021ApJ...910..150V},
	pages = {150},
}

@article{bryan_figure_2007,
	title = {Figure rotation of dark haloes in cold dark matter simulations},
	volume = {380},
	issn = {0035-8711},
	url = {https://ui.adsabs.harvard.edu/abs/2007MNRAS.380..657B},
	doi = {10.1111/j.1365-2966.2007.12096.x},
	urldate = {2022-10-18},
	journal = {Monthly Notices of the Royal Astronomical Society},
	author = {Bryan, S. E. and Cress, C. M.},
	month = sep,
	year = {2007},
	note = {ADS Bibcode: 2007MNRAS.380..657B},
	pages = {657--664},
}

@article{bailin_figure_2004,
	title = {Figure {Rotation} of {Cosmological} {Dark} {Matter} {Halos}},
	volume = {616},
	issn = {0004-637X, 1538-4357},
	url = {https://iopscience.iop.org/article/10.1086/424912},
	doi = {10.1086/424912},
	language = {en},
	number = {1},
	urldate = {2022-07-01},
	journal = {The Astrophysical Journal},
	author = {Bailin, Jeremy and Steinmetz, Matthias},
	month = nov,
	year = {2004},
	pages = {27--39},
}
\bibliographystyle{aasjournal}

\appendix 

\section{Idealized halo shape evolution}

\begin{figure}[h]
    \centering
    \includegraphics[width=0.8\linewidth]{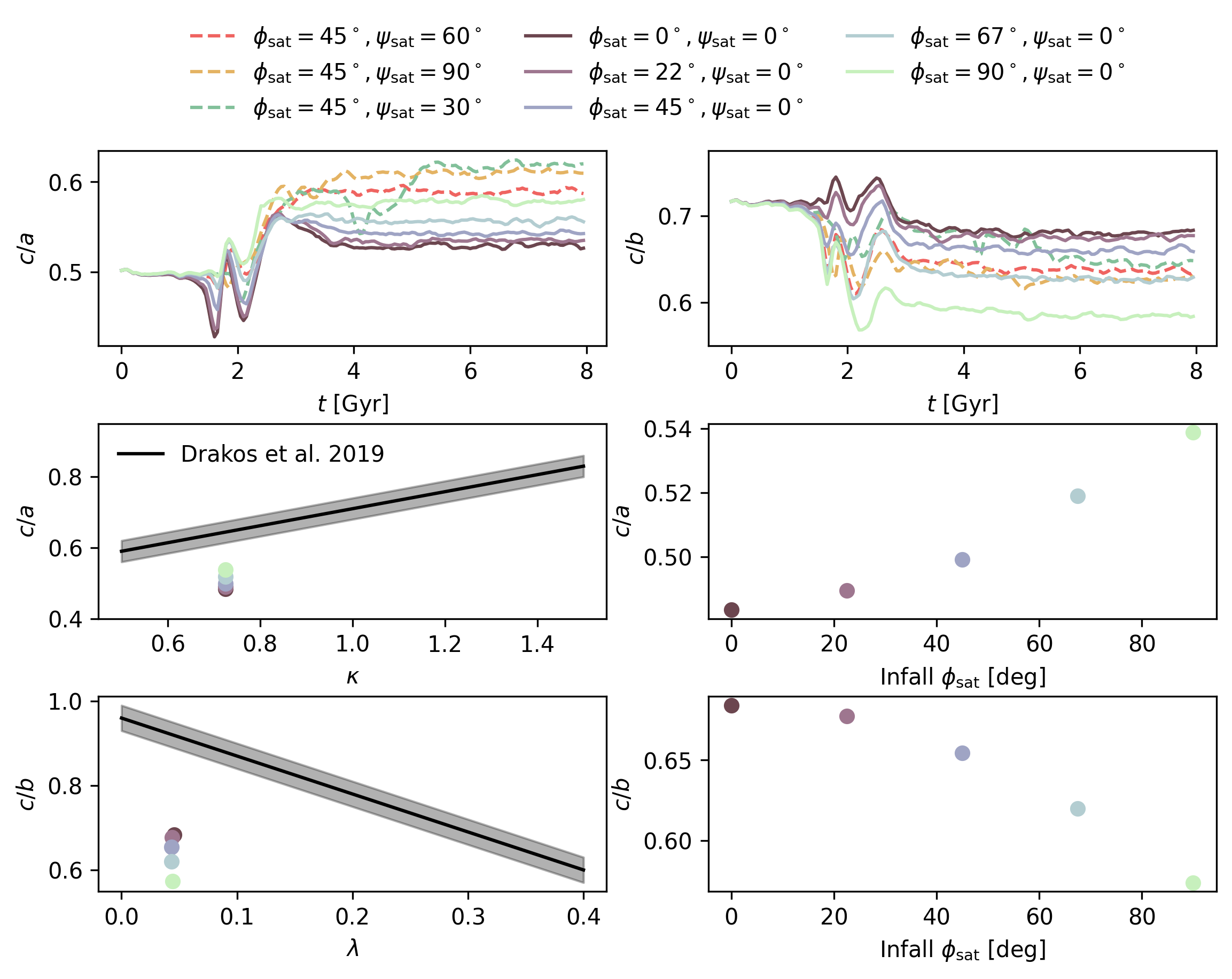}
    \caption{Halo shape evolution (top row) and remnant halo shape dependence on satellite orbital properties (middle and bottom rows). Dimensionless energy $\kappa$ and halo spin $\lambda$ are defined as in \cite{drakos_major_2019}, and the linear relations fit from their merger remnants and RMS scatter are plotted as the black line and grey shaded regions, respectively. We note that the axial ratios in the top row are measured with particles originally belonging to the host only, while the other two rows are measured using all particles to better compare with \cite{drakos_major_2019}.}
\label{fig:Infall_angle_shape_dependence_6_panel}
\end{figure}

\textedit{In each of our idealized simulations, the axial ratios remain stable after the merger concludes for the duration of the simulation (Figure \ref{fig:Infall_angle_shape_dependence_6_panel}, top row). We note that the merger in our model with $\phi_{\mathrm{sat}}=45^\circ,\psi_{\mathrm{sat}}=30^\circ$ does not conclude until $t\sim5$ Gyr.}

\textedit{Previous idealized simulations \citep[][, hereafter D19]{drakos_major_2019} have found that remnant halos from binary galaxy mergers have shapes which strongly depend on the orbit of the merging galaxies, with more radial mergers producing more prolate halos and more circular mergers producing oblate halos. We briefly compare the results of our simulations to those of D19, with the important distinction that we utilize a halo which is initially triaxial rather than spherical. In the middle and bottom panels of the left column of figure \ref{fig:Infall_angle_shape_dependence_6_panel} we show the minor:major axis ratio against dimensionless energy $\kappa$ and the minor:intermediate axis ratio against dimensionless spin $\lambda$ of the remnant halo, respectively, for our radial merger models. Each are measured at the final simulation snapshot. Overplotted on each of these panels are the linear fits obtained by D19 along with the RMS scatter of their simulations (grey shaded region). The remnant halos of our simulations do not fall along the linear relation and show an additional spread in shape which appears to be explained by the angle between the infall of the satellite and the host halo major axis (middle and bottom panels of the right column). We note that, because they simulate the merger of equal mass and spherical halos, we do not necessarily expect to match the results of D19. Our idealized simulations demonstrate that, in addition to orbital energy and eccentricity \citep{mcmillan_haloes_2007,drakos_major_2019}, remnant halo shapes are sensitive to the orientation of an initially triaxial host with respect to the infalling satellite.} 

\section{Derivation of substructure coefficient contribution}\label{sec:substructure_contribution_derivation}

We first assume that the substructure of interest is sufficiently compact and distant from the halo centroid that its density may be approximated as a delta function:

\begin{equation}
\begin{split}
    & \rho_{\text{sub}}(\mathbf{x}) = M_{\text{sub}}\delta^3(\mathbf{x}-\mathbf{x}_{\text{sub}}) \\ &= M_{\text{sub}}\frac{\delta(r-r_{\text{sub}})}{r^2}\frac{\delta(\theta-\theta_{\text{sub}})}{\sin \theta} \delta(\phi-\phi_{\text{sub}}).
\end{split}
\end{equation}
For the basis expansion of some density function, coefficients on a spherical shell at radius $a$ with thickness $\Delta a$ are calculated by convolving the density with the corresponding spherical harmonic function:
\begin{equation}
\begin{split}
    &\rho_{\ell m}(a)\Delta a = \oint d^2\mathbf{\Omega}Y_\ell^{m*}(\mathbf{\Omega})\rho(a,\mathbf{\Omega}) \\
    & = \int_0^\pi d\theta\sin\theta\int_0^{2\pi}d\phi Y_\ell^{m*}(\theta,\phi)\rho(a,\theta,\phi)\Delta a.
\end{split}
\end{equation}
In the case of a delta function density, the convolution simplifies to the value of the harmonic at the angular position of the substructure,
\begin{equation}
    \rho_{\ell m}^{\text{sub}}(a) = \frac{M_\text{sub}}{a^2}Y_\ell^m(\theta_{\text{sub}},\phi_{\text{sub}})\delta(a-r_\text{sub}),
\end{equation}
which is only non-zero when the satellite is contained by the shell. The potential coefficients $\Phi_{\ell m}$ are determined by integrating the internal and external contributions from the density \citep[see][chapter 2.4]{binney_galactic_2008},
\begin{equation}
    \Phi_{\ell m}(r) = - \frac{4\pi G}{2\ell+1}\left( r^{-\ell-1}\int_0^r a^{\ell+2} \rho_{\ell m}(a)da +\right.\left. r^\ell\int_r^\infty a^{1-\ell} \rho_{\ell m}(a) da \right).
\end{equation}
Substituting $\rho_{\ell m}^{\text{sub}}$ simplifies these integrals to give 
\begin{equation}
    \Phi_{\ell m}^{\text{sub}}(a)
= \frac{-4\pi G M_\text{sub}}{r_\text{sub}} \frac{1}{2\ell+1} Y_\ell^{m}(\theta_\text{sub},\phi_\text{sub}) \\ \times
\begin{cases}
 \left(\frac{r_\text{sub}}{a}\right)^{\ell+1} \text{ for }   r_\text{sub} < a \\
 \left(\frac{r_\text{sub}}{a}\right)^{-\ell} \text{ for }   a < r_\text{sub}
\end{cases}.
\end{equation}

The above derivation provides an estimate for the contribution to a given $\ell,m$ coefficient inflicted by some substructure that has not yet become phase-mixed, in the approximation that the substructure is reasonably compact. `Reasonably compact' here is applicable for substructures at large radii whose density distributions change relatively quickly in $\theta,\phi$ compared to the lower-order harmonics, and whose radial extent is smaller than the local shell width $\Delta a$. This is a good approximation for many subhalos, and may further serve to provide reasonable intuition for the contribution of the dynamical friction wake to the coefficient series. While we do not consider radial basis functions in our model, our approximation of the density as a delta function would allow for a similar estimation of its contribution to these coefficients as well.

\section{Stability of the idealized host halo}\label{sec:null_model_stability}

\begin{figure}
    \centering
    \includegraphics[width=.9\linewidth]{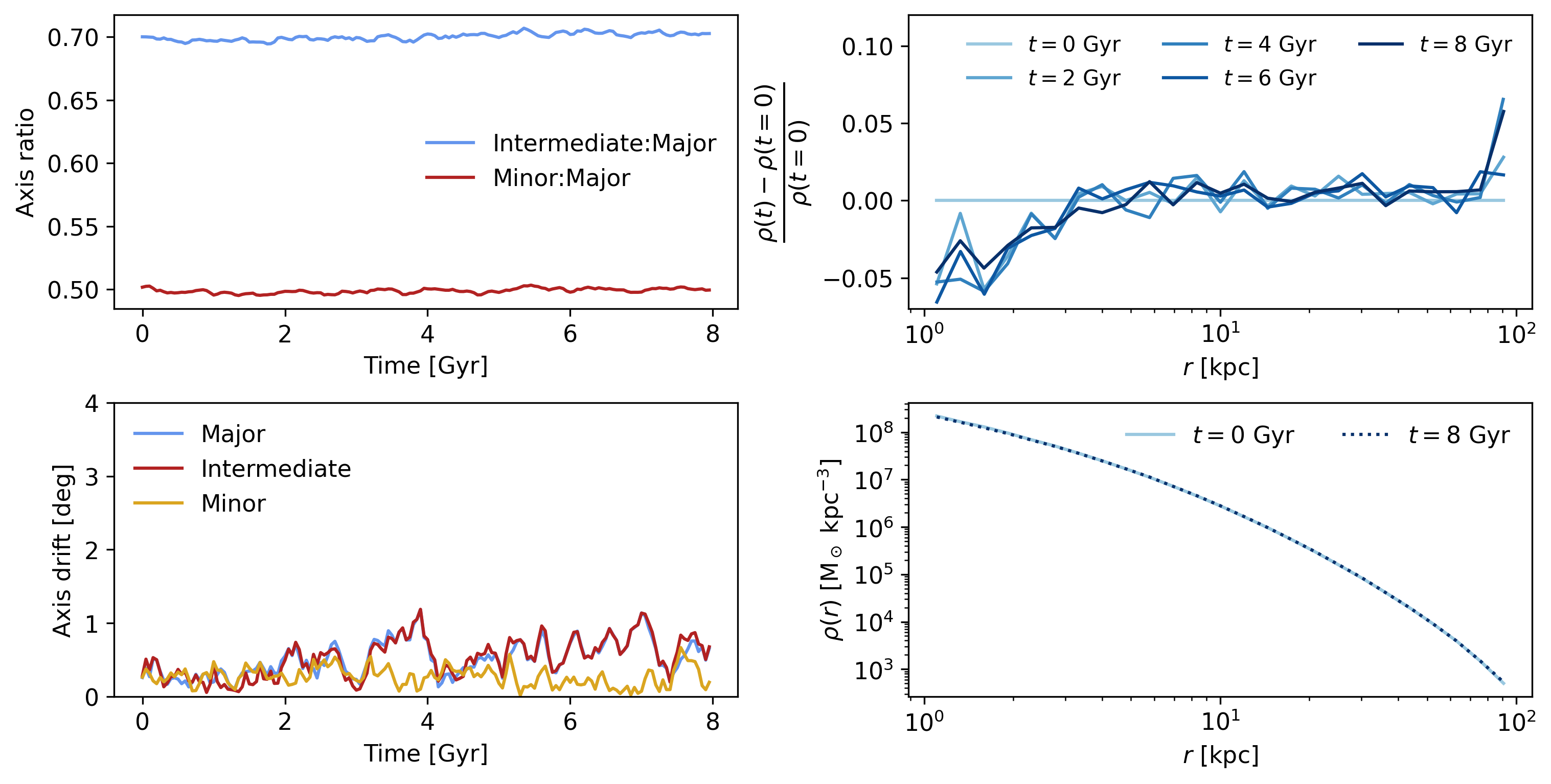}
    \caption{Evolution of the principal axes (left) and the density profile (right) of the halo used in our idealized simulations. To confirm system stability, we evolved the halo in isolation for 8 Gyr.}
\label{fig:null_model_axis_and_density_evolution}
\end{figure}

\textedit{To confirm the stability of our idealized halo, we evolve it in isolation for 8 Gyr and measure the evolution of its principal axis ratios and orientations, and density profile (see figure \ref{fig:null_model_axis_and_density_evolution}). The axis ratios (top left) are stable over the entire 8 Gyr time period, fluctuating by $\lesssim1\%$. Similarly, the orientations of the principal axes (bottom left) drift by less than a degree over the same time period. The density profile (top right) in the inner $\sim2$ kpc of the halo is lowered by $\lesssim5\%$ early in the simulation but fluctuates by $1\%$ or less through the majority of the halo over the course of the simulation. The density profile (bottom right) remains stable over the 8 Gyr simulation with our Schwarzschild model initial conditions and time stepping routine. The apparent evolution in the density structure of the inner $\sim2$ kpc is likely due to our gravitational softening procedure, which modifies the interparticle forces and hence the gravitational potential from that in used to create the ICs on scales comparable to the affected region. The effect is both small ($\sim5\%$) and restricted to the inner $\sim2$ kpc, and we therefore consider it to be acceptable.}

\section{Resolution convergence}\label{sec:resolution_convergence}

\begin{figure}[h]
    \centering
    \includegraphics[width=0.8\linewidth]{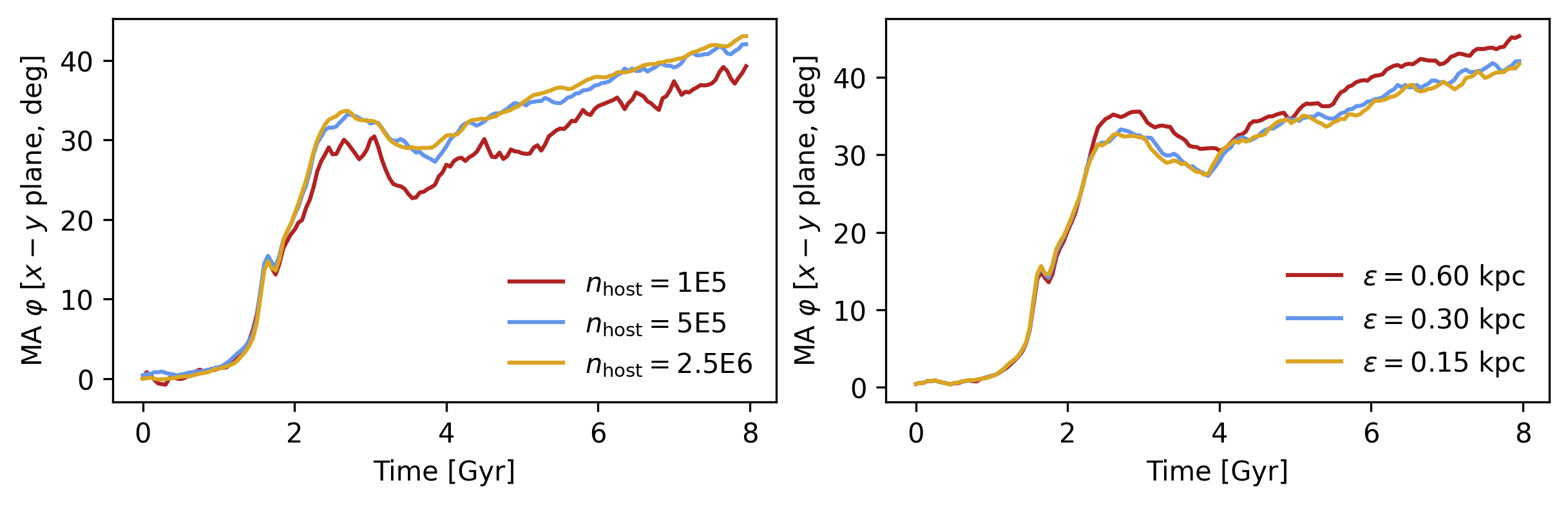}
    \caption{Resolution convergence tests for our idealized simulations. We run two suites scaling the number of particles by a factor of 5 higher and lower (left) and the softening length by a factor of 2 higher and lower (right) than their fiducial values.}
    \label{fig:Resolution_convergence_tests}
\end{figure}

\textedit{We briefly check the resolution convergence of our simulations in both particle number and softening length (see figure \ref{fig:Resolution_convergence_tests}, left and right panels respectively) for our radial infall simulation with $\phi_{\mathrm{sat}}=45^\circ$. Utilizing 5 times the number particles from our main run simulations (yellow curve) yields identical results to our fiducial resolution (blue curve). Reducing the particle count by a factor of 5 (red curve) diminishes the swinging of the major axis during the transient phase but does not seem to change the post-merger tumbling pattern speed. Similarly, reducing the softening length by a factor of two (bottom panel, yellow curve) does not appreciably change the results from our fiducial softening length (blue curve), while increasing the softening by a factor of two (red curve) increases the swinging of the major axis during the transient phase without significantly affecting the pattern speed post-merger. When varying the softening length, we inversely scale the simulation timestep so that close interactions remain temporally resolved. }

\end{document}